%% file: ms.tex
\documentclass{emulateapj}

\usepackage{amsmath}

\slugcomment{Accepted by the Astrophyiscal Journal: 19 December 2010}

\shorttitle{Cluster AGNs}
\shortauthors{Atlee et al.}

\begin{document}
\title{A Multi-Wavelength Study of Low Redshift Clusters of Galaxies I. Comparison of X-ray and Mid-Infrared Selected AGNs}
\author{David W. Atlee, Paul Martini, Roberto J. Assef}
\affil{Department of Astronomy, The Ohio State University, 140 W. $18^{th}$ Ave. Columbus, OH 43210, USA}
\email{atlee@astronomy.ohio-state.edu}
\and
\author{Daniel D. Kelson, John S. Mulchaey}
\affil{The Observatories of the Carnegie Institution of Science, 813 Santa Barbara Street, Pasadena, CA 91101, USA}

\begin{abstract}
Clusters of galaxies have long been used as laboratories for the study of galaxy
evolution, but despite intense, recent interest in feedback between AGNs
and their hosts, the impact of environment on these relationships
remains poorly constrained.
We present results from a study of AGNs and their host galaxies
found in low-redshift galaxy clusters.
We fit model spectral energy distributions (SEDs)
to the combined visible and mid-infrared (MIR) photometry of cluster
members and use these model SEDs to determine stellar masses 
and star-formation rates (SFRs).
We identify two populations of AGNs, the first based on their X-ray
luminosities (X-ray AGNs) and the second based on the presence of a
significant AGN component in their model SEDs (IR AGNs).  
We find that the two AGN populations
are nearly disjoint; only 8 out of 44 AGNs are identified with
both techniques.  We further find that
IR AGNs are hosted by galaxies with similar masses and SFRs
but higher specific SFRs (sSFRs) than X-ray AGN hosts.
The relationship between AGN accretion and host star-formation
in cluster AGN hosts shows no significant difference compared
to the relationship between field AGNs and their hosts.
The projected radial distributions of both AGN populations are
consistent with the distribution of other cluster members.
We argue that the apparent dichotomy between X-ray and IR AGNs 
can be understood as a combination of differing extinction due to cold gas
in the host galaxies of the two classes of AGNs and the
presence of weak star-formation in X-ray AGN hosts.
\end{abstract}

\keywords{galaxies:active, galaxies:clusters:general, infrared radiation,
X-rays}

\section{Introduction}\label{secIntro}
Galaxy formation and evolution has long been a subject of considerable
interest, with early work dedicated to exploring the physical
processes responsible for star-formation \citep{whip46}, explaining
the genesis of the Milky Way \citep{egge62}, and examining the
evolution of galaxies in clusters \citep{spit51}.
Models for the evolution of galaxies in clusters gained strong observational
constraints with the discovery of an apparent evolutionary sequence
among local clusters \citep{oeml74}.  The discovery that
the fraction of blue, spiral galaxies in relaxed galaxy clusters increases
from $z=0$ to $z\approx0.4$ quickly followed \citep{butc78,butc84}.
The dearth of spiral galaxies in
the high-density regions at the centers of 
galaxy clusters is known as the morphology-density relation
\citep{dres80,post84,dres97,post05}. 
This relation places additional, strong constraints on evolutionary
models for cluster galaxies.
That star-forming galaxies are also rare in the 
centers of clusters had been previously
suggested by the results of \citet{oste60} and was subsequently
observed in other work \citep{gisl78,dres85}.
The impact of environment on the frequency and intensity of star-formation
at a wide variety of density scales has been measured using numerous
visible \citep{abra96,balo97,kauf04,pogg06,pogg08,vdLi10}
and mid-infrared (MIR;
\citealt{sain08,bai09}) diagnostics.  Star-forming galaxies
are consistently found to be
more common and to have higher star-formation rates (SFRs) 
in lower density environments and at higher redshift
\citep{kauf04,pogg06,pogg08}.

The observed trends in star-formation with environment are
usually attributed to variations in the sizes of gas reservoirs, either
the existing cold gas or the hot gas that can
cool to replenish the cold gas as it is consumed.
Given that AGNs also consume cold gas to fuel their luminosity,
similar patterns might be expected among AGNs.
Indeed, recent work reveals strong dependencies of the luminosities and types
of AGNs on environment (e.g.\ \citealt{kauf04,pope06,cons08,mont09})
for AGNs selected via visible-wavelength 
emission-line diagnostics.  Von der Linden et al.\ (2010)
find fewer ``weak AGNs'' (primarily LINERS)
among red sequence galaxies near the centers
of clusters compared to the field, but they find no corresponding
dependence among blue galaxies.  Intriguingly, while \citet{mont09}
independently report a decline in the fraction of low-luminosity AGNs toward
the centers of low-redshift clusters, they find an {\it increase} in
the fraction of LINERs in higher density environments.  The 
difference is likely a result of evolution.  \citet{mont09} found
qualitatively different behavior between their main $z\sim1$ sample
and the result produced when they applied their analysis to SDSS clusters.
These results indicate that the variation
of galaxy properties with local environment may influence
the types of AGNs observed and that evolution in the relationship
between some AGN classes and their host galaxies is important.
Understanding the environmental mechanism that
transforms star-forming galaxies into passive galaxies
in clusters may help relate gas reservoirs in cluster galaxies to 
galaxy evolution as well as to AGN feeding and feedback.

Several mechanisms to cause the transformation from star-forming to
passive galaxies have been proposed.
These include ram-pressure stripping of cold gas 
\citep{gunn72,quil00,roed05}, strangulation \citep{lars80,balo00,kawa08,mcca08}
and galaxy harassment \citep{moor96,moor98,lake98}.
Each mechanism operates on a different characteristic timescale and has its
greatest impact on galaxies of different masses and at different
radii.  In principle, the transition of galaxy populations from star-forming
to passive as a function of environment can probe the relative importance
of these processes.
However, such approaches suffer from practical difficulties.
For example, \citet{bai09} argue that the similarity
of the $24\mu m$ luminosity functions observed in galaxy clusters
and in the field suggests that the transition from star-formation to
quiescence
must be rapid, which implies that ram pressure stripping must be the
dominant mechanism.  Von der Linden et al.\ (2010),
by contrast, find a significant
trend of increasing star-formation with radius up to $5 R_{200}$ from cluster
centers.  They conclude that preprocessing at the group scale is
important, which is inconsistent with ram pressure stripping as the
driver of the SFR-density relation.
\citet{pate09} find a similar trend for increasing average SFR with
decreasing local density down to group-scale densities
($\Sigma_{gal}\approx1.0\ {\rm Mpc^{-2}}$) near RX J0152.7-1357 ($z=0.83$).
The importance of preprocessing in group-scale 
environments reported by these authors 
suggests that strangulation rather than ram pressure stripping
drives the SFR-density relation.
The starkly different conclusions reached by \citet{bai09} compared to
\citet{pate09} and \citet{vdLi10},
despite their common use of star-forming galaxies
to examine the influence of environment,
highlight the difficulties inherent in such studies.

Attempts to distinguish between various environmental
processes become still more difficult with cluster samples
that span a wide range in redshifts.  The
epoch of cluster assembly ($0\leq z\lesssim 1.5$, e.g.\ \citealt{berr09})
coincides with the epoch of 
rapidly declining star formation 
(e.g.\ \citealt{mada98,hopk06}) and AGN activity 
(e.g.\ \citealt{shav96,boyl98,shan09}),
which makes it difficult to disentangle rapid environmental effects
from the global reduction in the amount of available cold gas.
\citet{dres83} found early evidence for an increase
in AGN activity with redshift, and the Butcher-Oemler effect had already
provided evidence for a corresponding increase in SFRs.  In the
last decade, the proliferation of observations of high-redshift galaxy
clusters at X-ray, visible and infrared wavelengths
has yielded similar trends in the fraction of both AGNs
\citep{east07,mart09} and 
star-forming galaxies \citep{pogg06,pogg08,sain08,hain09}
identified using a variety of methods.  These newer results have also
examined cluster members confirmed from spectroscopic redshifts
rather than relying solely on statistical excesses
in cluster fields, which permits more detailed study of the relationships
between galaxies and their parent clusters.

The wide variety of AGN selection techniques employed in more recent studies
represents an important step forward in understanding the dependence of AGNs
on environment.  
Several recent papers have used X-rays to study the
frequency and distribution of AGNs in galaxy clusters 
(\citealt{mart06}, henceforth M06; \citealt{mart07,siva08,arno09,hart09}) 
and their evolution with redshift \citep{east07,mart09}.
\citet{mart09} found that the AGN fraction among cluster members increases
with decreasing local density and increases dramatically
($f_{AGN}\propto(1+z)^{5.3\pm1.7}$) with redshift.  They also found
that X-ray identification produces a much larger AGN sample than
visible-wavelength emission line diagnostics: only 4 of the
35 X-ray sources identified as AGNs by M06 would be classified as 
AGNs from their visible-wavelength emission lines.  Similar 
results have been found
when comparing radio, X-ray and mid-IR AGN selection techniques for field AGNs
(e.g.\ \citealt{hick09}).

The different AGN selection techniques identify different
AGN populations and suffer from distinctive selection biases.
Both X-ray and visible-wavelength techniques
can miss AGNs due to absorption, either in the host galaxy or in the
AGN itself; however, X-ray selection can find lower luminosity AGNs and AGNs
behind larger absorbing columns compared to emission line selection.
Mid-infrared selection techniques suffer from relatively
poor angular resolution, so they
are mainly sensitive to AGNs that outshine their host galaxies in
the band(s) used to perform the AGN selection.  The X-ray and visible
techniques can also be contaminated by emission from the host galaxy.
While the identification of X-ray sources with 
$L_{\rm X}>10^{42}\ {\rm erg\ s^{-1}}$  as AGNs is
unambiguous, X-ray luminosities in the 
$10^{40}$--$10^{42}\ {\rm erg\ s^{-1}}$ range can be produced by 
low-mass X-ray binaries (LMXBs), high-mass X-ray binaries (HMXBs), and
thermal emission from hot gas.  Both visible-wavelength and MIR
indicators are subject to contamination from
young stars, which produce emission lines and heat dust near star-forming
regions until it emits in the MIR.
Even the interpretation of the well-established
Baldwin-Phillips-Terlevich diagram \citep{bald81}
can be controversial in the transition
region between star-forming galaxies and AGNs.

These difficulties motivate the use of multiple techniques
to obtain a complete census of AGN and to correctly identify
potential imposters.  In this paper, we extend the work of
Martini et al.\ (2006, 2007) by supplementing their
X-ray imaging and visible-wavelength photometry 
with MIR observations from the Spitzer
Space Telescope.  We use these data to select AGNs 
independent of their X-ray emission.  We also measure
the properties of AGN host galaxies by fitting their visible to MIR
spectral energy distributions (SEDs).
We discuss our visible and MIR data reduction and photometry
in Section \ref{secObs}.
Section \ref{secMethods} details our techniques for identifying AGNs
and measuring galaxy properties, and we describe the results 
in Section \ref{secResults}.  We discuss the 
implications for the relationship between AGNs and their host
galaxies in Section \ref{secDiscuss}.
Throughout this paper we use the WMAP 5-year cosmology---a $\Lambda$CDM 
universe with $\Omega_{m}=0.26$,
$\Omega_{\Lambda}=0.74$ and $h=0.72$ \citep{dunk09}.

\section{Observations \& Data Reduction}\label{secObs}
We obtained MIR observations with the {\it Spitzer} Space Telescope of the 
X-ray sources identified as members of 8 low-redshift galaxy clusters
by M06.
The initial reduction of the Spitzer imaging is described in Section 
\ref{secSpitzer}.  Visible wavelength photometry of these
clusters were obtained at the 2.5m du Pont telescope at Las Campanas
by M06.  We provide a brief summary of these data in Section \ref{secOptical};
further details are provided by M06.  We then discuss the corrections
for Galactic extinction and for instrumental effects in Section \ref{secCorr}.

\subsection{Spitzer Reduction}\label{secSpitzer}
We obtained mid-infrared (MIR) observations from the {\it Spitzer}
Space Telescope using the IRAC ($\lambda_{eff}=3.6,4.5,5.8,8.0\ \mu m$;
\citealt{fazi04}) and MIPS ($\lambda_{eff}=24$; \citealt{riek04}) instruments
from {\it Spitzer} program 50096 (P.I.\ Martini).
Observations were carried out between 2008 November 1 and 2009 April 22.
{\it Spitzer} pointings were chosen to image the X-ray point sources
in 8 low-redshift galaxy clusters identified by M06.  We supplemented
these observations with data from the {\it Spitzer} archive for Abell 1689
and AC 114.

{\it Spitzer}'s cryogen ran out before the MIPS
observations of three clusters (Abell 644, Abell 1689
and MS 1008.1-1224) were carried out.  In one of these clusters 
(Abell 1689) we extended our coverage to $24\mu m$ using observations
from the {\it Spitzer} archive,
leaving two clusters with no usable MIPS observations.
The Astronomical Observation Request (AOR) numbers used to construct
the MIR mosaic images of each cluster
are listed in Table 1, along
with the corresponding $3\sigma$ observed-frame 
luminosity limits at both $8$ and
$24\mu m$.
These limits are approximate because the image depth
varies across the mosaics due the changing number of
overlapping pointings.  Quoted limits correspond to
areas with ``full coverage'' but without overlap from adjacent pointings.

The raw {\it Spitzer} data are reduced by an automated pipeline before they
are delivered to the user, but artifacts inevitably remain in the
calibrated (BCD) images.
Preliminary artifact mitigation for the IRAC images was performed using the 
IRAC artifact mitigation tool by Sean 
Carey\footnote{http://spider.ipac.caltech.edu/staff/carey/irac\_artifacts/}.
We inspected each corrected image after this step and determined whether
the image was immediately usable, if additional corrections were required,
or if it simply had too many remaining artifacts to be reliably corrected.
The latter class primarily included images with extremely bright
stars that caused artifacts too severe to be corrected.
Where appropriate, additional corrections were applied using the
muxstripe\footnote{http://ssc.spitzer.caltech.edu/dataanalysistools/tools/contributed/\\irac/automuxstripe/}
and jailbar\footnote{http://ssc.spitzer.caltech.edu/dataanalysistools/tools/contributed/\\irac/jailbar/}
correctors by
Jason Surace and the column pull-down 
corrector\footnote{http://ssc.spitzer.caltech.edu/dataanalysistools/tools/contributed/\\irac/cpc/}
by Leonidas Moustakas.  Artifacts in the MIPS images were 
removed by applying a flatfield correction algorithm packaged with
the {\it Spitzer} mosaic software,
(MOPEX\footnote{http://ssc.spitzer.caltech.edu/dataanalysistools/tools/mopex/}),
as described on the Spitzer Science Center (SSC)
website\footnote{http://ssc.spitzer.caltech.edu/dataanalysistools/cookbook/\\23/\#\_Toc256425880}.

Mosaic images for both IRAC and MIPS were constructed from the
artifact-corrected images using MOPEX.  Aperture photometry was 
extracted from the resulting mosaics using the {\it apphot} package
in IRAF.
We converted the measured fluxes to magnitudes in the Vega system after
the photometric corrections described in Section
\ref{secCorr} had been applied.  All
magnitudes quoted in this work, both visible and MIR, are calculated with
respect to the Vega standard.
The photometric apertures used by {\it apphot}
were chosen to enclose a region of
approximately 10 kpc projected radius at the redshift of each cluster.
These large apertures yielded reduced S/N, but 
most cluster members were sufficiently bright that the uncertainties 
on the measured fluxes were
dominated by systematic errors (5\%) in the zero-point calibration, 
except at $24\mu m$.  The use of large photometric apertures also 
allowed galaxies
to be treated as point sources for the purpose of computing aperture
corrections, as recommended by the SSC.  
A smaller aperture could improve the S/N,
but this gain would be outweighed by the systematic uncertainty
introduced by the aperture corrections for the resulting flux measurements,
as aperture corrections for IRAC extended sources remain highly uncertain
(IRAC Instrument Handbook\footnote{http://ssc.spitzer.caltech.edu/irac/iracinstrumenthandbook/\\IRAC\_Instrument\_Handbook.pdf}).

\subsection{Visible Photometry}\label{secOptical}
All 8 clusters in our sample have $B$-, $V$- and $R$-band imaging,
and 4 of the 8 have $I$-band imaging.
We extracted separate source catalogs for each of these
bands using Source Extractor (SExtractor, \citealt{bert96}) and 
merged the catalogs using the $R$-band image as the reference image for
astrometry and total (Kron) magnitudes. 
We correct from aperture to total magnitudes without
altering the colors from the aperture photometry
by applying the $R$-band aperture corrections to all bands,
\begin{equation}\label{eqAperCorr}
m_{Kron} = m_{Ap} - (R_{Ap}-R_{Kron})
\end{equation}
where $m_{Ap}$ and $m_{Kron}$ are the aperture and Kron-like magnitudes,
respectively, for the band being corrected.  Rather than taking the
published photometry from M06, we used the redshift-dependent 
apertures assigned to each cluster as described in Section
\ref{secSpitzer}.  This maintains consistency with our IRAC photometry and
results in relatively small aperture corrections,
typically $\sim0.1$ mag.

SExtractor returns $R$-band positions 
that are good to within a fraction of an
arcsecond.  However, the positions of sources in IRAC and MIPS images are
less precise due to the poorer angular resolution and larger
pixel sizes in these bands.  We selected the best astrometric matches to
each Spitzer source from the objects identified by SExtractor within a
specified search radius, $\theta$.  To determine the best value of
$\theta$, we scrambled the RA of SExtractor sources and determined how
many Spitzer sources were matched to a scrambled galaxy
as a function of $\theta$.  We found
the best balance between purity and completeness for 
$\theta\approx1\farcs{25}$.  This search radius yielded spurious
matches for less than
2\% of objects.  The actual contamination of our catalog will be much lower,
because a Spitzer object with a spurious match will
usually be better matched to its ``true'' counterpart,
which has a median match distance $d=0\farcs{4}$.  The images
used to perform the matching do not suffer from substantial confusion, even
in the cluster centers, so erroneous photometry due to overlapping
sources is unlikely to present a problem.
Further details of the visible image reduction were described by M06.

\subsection{Photometric Corrections}\label{secCorr}
We estimated the Galactic reddening toward each of the 8 clusters
in our sample 
from the dust map of \citet{schl98} and calculated
extinction corrections assuming $R_{V}=3.1$ and the \citet{card89}
reddening law.
The resolution of the \citet{schl98} dust map requires us to use a
common extinction correction for all cluster members.
However, Galactic
cirrus is apparent in some of our images, so this assumption is not always
appropriate.  This leads to additional uncertainty associated
with the extinction corrections, but the total (visual) extinction toward
our clusters is typically less than 0.1 mags.  The
associated uncertainties are therefore small.
For the clusters with the
highest extinctions (Abell 2104 and 2163, with $A_{V}=0.73$ and 1.1, 
respectively), variations in extinction
across the cluster represent an
important source of systematic uncertainty.  We account for this
by adopting a 10\% uncertainty in all extinction corrections and
propagating this uncertainty to the corrected magnitudes.  In Abell 2163,
for example, this corresponds to an uncertainty of 0.11 mags in the 
dereddened $V$-band magnitude.

The raw fluxes measured from the MIR mosaics must be corrected for various
instrumental effects, including aperture, array-location and color corrections,
as described in the IRAC
and MIPS\footnote{http://ssc.spitzer.caltech.edu/mips/mipsinstrumenthandbook/\\MIPS\_Instrument\_Handbook.pdf}
Instrument Handbooks.  Aperture
corrections are, in principle, required for all observations.  In
practice, even our smallest apertures ($\sim7''$) are large enough that
aperture corrections to visible-wavelength point sources are negligible.
For MIR point sources, this is not the case.
We apply aperture corrections from the IRAC Instrument Handbook
appropriate for our redshift-dependent photometric apertures to the
IRAC photometry.
These corrections are not strictly appropriate due to the extended nature
of our sources; however, we have chosen apertures that are large
compared to the sources ($\sim 3\times$ larger than 
the FWHM of the largest galaxies, see Section \ref{secSpitzer}).  We 
therefore apply aperture
corrections appropriate for point sources.  

We determined aperture corrections appropriate for our MIPS images
by averaging a theoretical point-source response function (PRF) from 
STinyTim\footnote{http://ssc.spitzer.caltech.edu/dataanalysistools/tools/\\contributed/general/stinytim/}
with three bright, isolated point sources in the Abell 3125
and Abell 2104 mosaics.
The PRFs of sources from the different clusters agree with one another and with
the theoretical PRF to within a few percent over the range of aperture
sizes relevant for our MIPS photometry.
The dispersion between the individual PRFs at fixed aperture
size provides an estimate of the uncertainty on the corrections and is
included in the $24\mu m$ error budget.
The MIPS images of the other clusters lack bright, isolated
points sources with which to make a similar measurement, so we assume that
the PRF appropriate for Abell 3125 and Abell 2104
gives reasonable aperture corrections for
all clusters.  This introduces some systematic error in our derived $24\mu m$
fluxes, but the agreement of the observed PRFs of point-sources
identified in 
Abell 3125 and Abell 2104 with the theoretical PRF indicates that this 
uncertainty is small.

The flatfield corrections applied to IRAC images by the automated
image reduction pipeline are based on observations of the zodiacal background
light, which is uniform on the scale of the IRAC field
of view.  It is also extremely red compared to any normal astrophysical
source.  The combination of scattered light due to the
extended nature of the source and the color of the source illuminating
the detector for the flatfield images results in different gains for
point-sources and extended sources.  It also requires 
an effective bandpass
correction that varies with position on the detector.  These
effects can be corrected by applying a standard array-location correction image
to a single IRAC image.  For a mosaic, the magnitude of the required
correction is significantly
reduced by adding dithered images with different corrections
at a given position on the sky.  However,
the residual effect can be a few percent or more depending on the number of
overlapping IRAC pointings.  We construct an array-location correction
mosaic by co-adding the correction image for a single IRAC pointing
shifted to the positions of each dithered image in the
science mosaic.  We measure the required array-location corrections in the
same apertures used to measure the IRAC fluxes.

The {\it Spitzer} image reduction pipeline assumes a flat power-law SED
to convert electrons to incident fluxes.
Astrophysical sources typically do not 
show flat SEDs and therefore require color
corrections to determine the true flux at the effective wavelength
of a given band.  This is especially important in star-forming galaxies,
which show strong polycyclic aromatic hydrocarbon (PAH) emission features
at 6.2 and $7.7\mu m$ \citep{smit07}.
We determine color corrections to the measured fluxes from
model SEDs (Section \ref{secModelSeds}).  We compute preliminary model
SEDs for each cluster member from the photometry with all other
corrections applied.  We then integrate the model SED across the various MIR
bandpasses and determine the appropriate color corrections following the
procedures outlined in the instrument handbooks.  The
color correction, $K$, applied to an IRAC source is given by,
\begin{equation}\label{eqColorCorr}
K=\frac{\int(F_{\nu}/F_{\nu_{0}})(\nu/\nu_{0})^{-1}R_{\nu} \mathrm{d}\nu}{\int(\nu/\nu_{0})^{-2}R_{\nu}\mathrm{d}\nu}
\end{equation}
where $F_{\nu}$ is the model spectrum and $R_{\nu}$ is the response 
function of the detector in the appropriate channel.  The formalism
for MIPS color 
corrections is similar but slightly more complicated;
we refer interested readers to
Section 3.7.4 of the MIPS Instrument Handbook.  Optical and MIR photometry
for each cluster member after all relevant corrections 
have been applied are listed in Table 2.

\section{Methods}\label{secMethods}
We wish to identify cluster members hosting AGNs, determine the
AGN luminosities, examine the properties of AGN host galaxies,
and determine whether they differ in any appreciable way from ``normal''
cluster galaxies or from their counterparts in the field.
This requires that we distinguish cluster members from
foreground and background galaxies, fit model
SEDs to the member photometry, and measure the rest-frame properties
of the AGN host galaxies.  We describe the model SEDs
in Section \ref{secModelSeds}.  Using these models, we calculate
K-corrections to the measured fluxes, estimate stellar masses and SFRs for
cluster member galaxies, and identify AGNs.

We use redshifts reported in \citet{mart07} or extracted from 
the NASA Extragalactic Database\footnote{http://nedwww.ipac.caltech.edu/}
to identify members of the galaxy clusters in our sample.
We define a galaxy to be a cluster member if it falls within a 
circular field with radius, 
\begin{align}\label{eqR200}
&R<R_{200}=1.7h^{-1}\ Mpc \biggl[\frac{\sigma}{1000\ {\rm km\ s^{-1}}}\biggr] \nonumber \\
&[(1+z)^{3}\Omega_{m}+\Omega_{\Lambda}]^{-1/2}
\end{align}
where $\sigma$ is the cluster's velocity dispersion \citep{treu03}.
We also require that members have spectroscopic redshifts within the
$\pm3\sigma$ redshift limits prescribed in Table 1 of \citet{mart07},
which were established using the biweight 
velocity dispersion estimator of \citet{beer90}.  This criterion yields
a sample of 1165 cluster member galaxies.  We eliminate many of these
galaxies from our sample due to either limited photometric coverage
or, in a few instances, because the spectroscopic redshifts in
the literature are clearly in disagreement with the photometric
redshifts obtained from the SED fits (Section \ref{secModelSeds}).
The final sample of ``good'' cluster members, those galaxies with
detections in at least 5 bands and with apparently reliable
spectroscopic redshifts, contains 488 galaxies.

\subsection{Model SEDs}\label{secModelSeds}
Assef et al.\ (2010; hereafter A10) constructed empirical SED templates 
that can be used to
determine photometric redshifts and K-corrections for galaxies
and AGNs over a wide
range of redshifts.  The A10 templates include three galaxy
templates (elliptical, spiral, and starburst or irregular)
and a single AGN template,
which can be subjected to variable intrinsic reddening.
These templates were derived
empirically across a long wavelength baseline ($0.03$--$30\mu m$),
using 14448 apparently ``pure'' galaxies and 5347 objects showing
AGN signatures.
We fit two independent model SEDs to the photometry of each cluster member 
using the published codes of A10.
The first model included only the
three galaxy templates, while the second also included an AGN component.
The $\chi^{2}$ differences between the two fits
can be used to identify AGNs (Section \ref{secAgn}).
Model SEDs for the M06 X-ray point sources included in our
sample of ``good'' galaxies are shown in Figure \ref{figXrayFits}.
AGNs identified from their SED fits, but which have no
X-ray counterparts, are shown in Figure \ref{figIrFits}.
The fits to the X-ray point sources are representative of the fit quality
returned for all cluster members, while the fits to photometrically-identified
AGNs are, on average, poorer.

\begin{figure*}
\epsscale{1.0}
\plotone{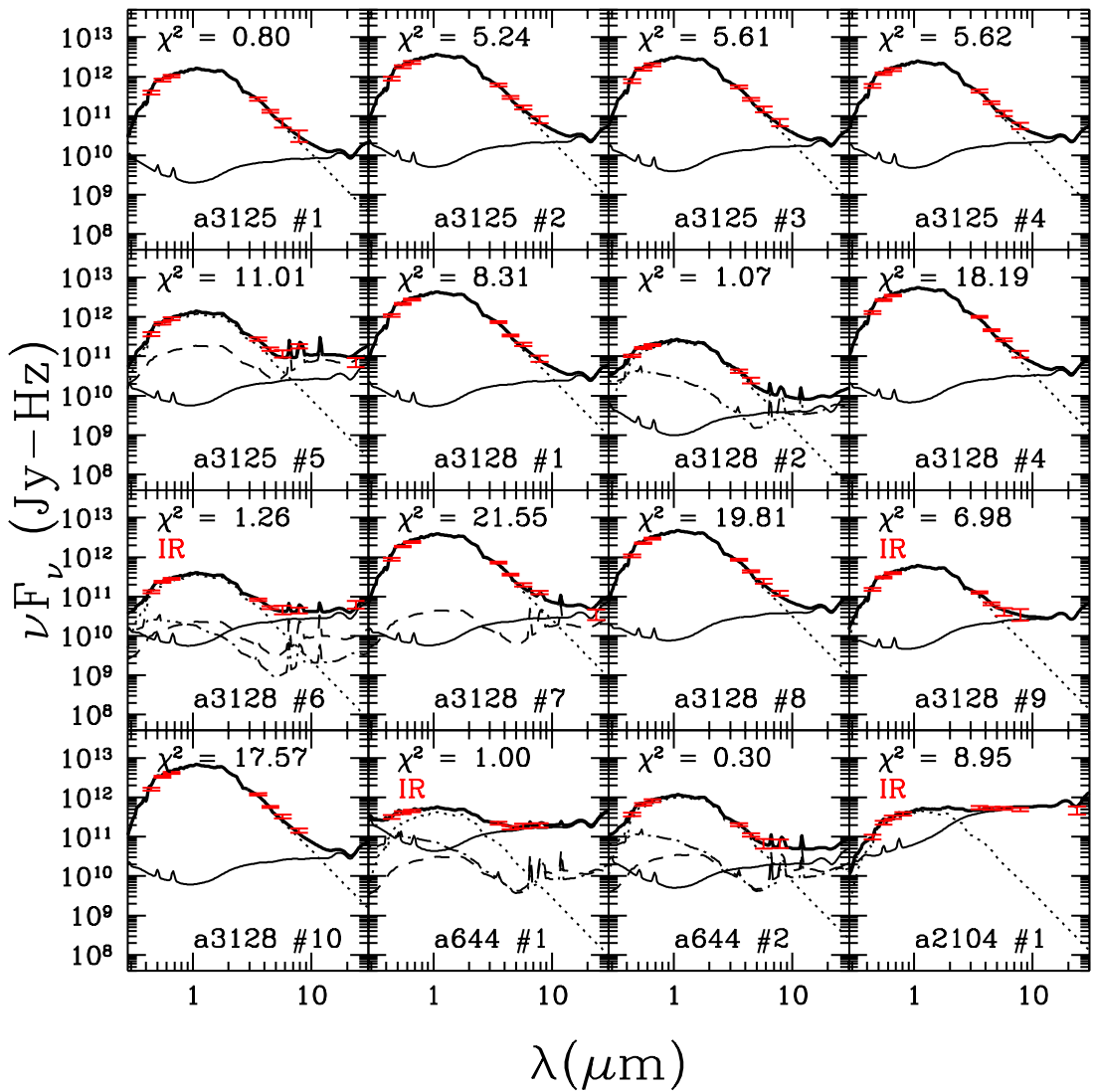}
\caption{Model SEDs for galaxies hosting M06 X-ray point sources.
Bands shown are, in order of wavelength,
$B$, $V$, $R$, $I$, $[3.6]$, $[4.5]$, $[5.8]$, $[8.0]$ and $[24.0]$.
The panels are labeled with the names assigned by 
M06 in their Table 4.  Objects also identified as AGNs from their
SED fitting are labeled ``IR.'' 
The heavy lines show the total model SED, while the {\it solid},
{\it dotted}, {\it dashed} and {\it dot-dashed} lines show the A10
{\it AGN, elliptical, spiral and irregular} templates, respectively.
Not all components appear in all panels.  
See Section \ref{secModelSeds} for further details.
\label{figXrayFits}}
\end{figure*}

\begin{figure*}
\addtocounter{figure}{-1}
\plotone{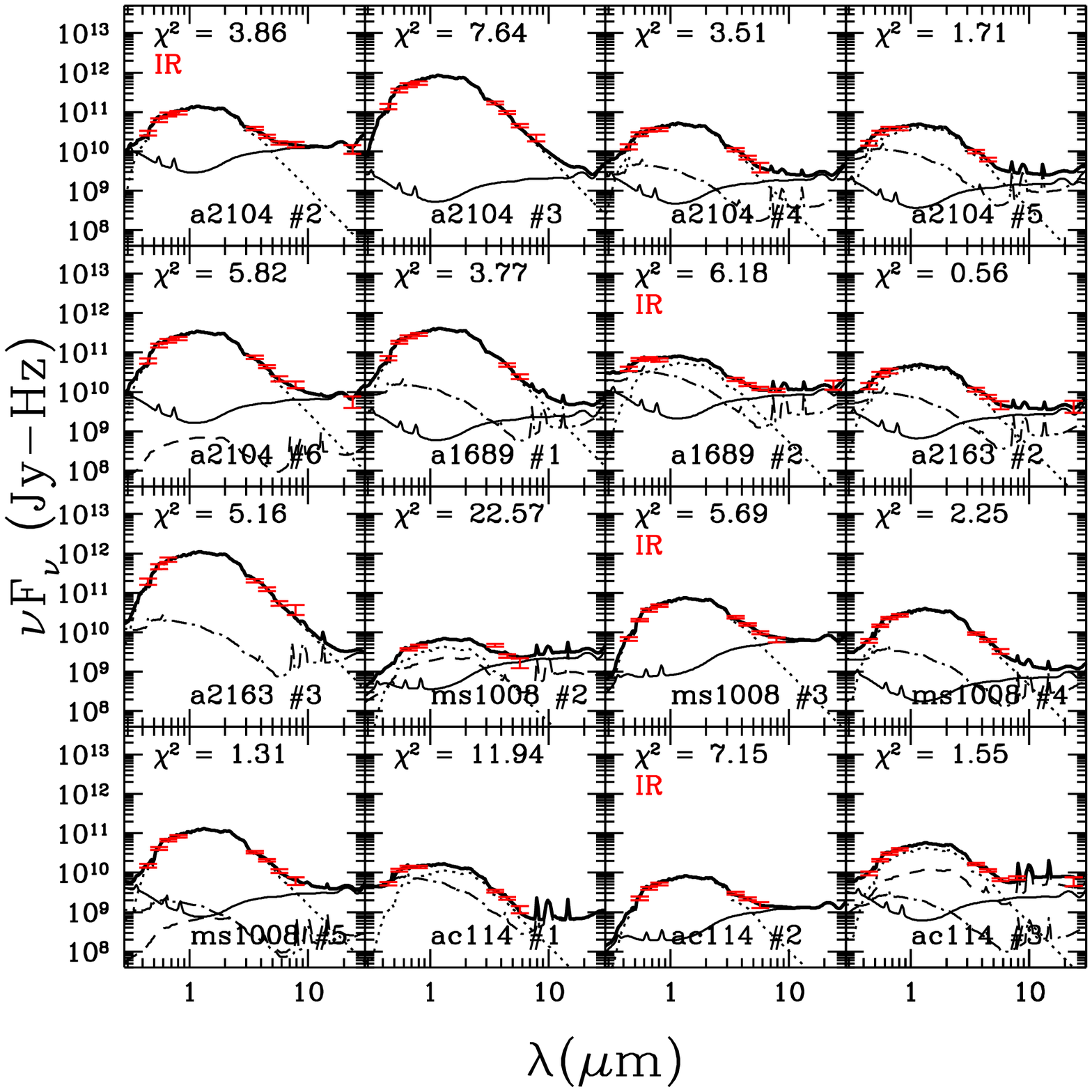}
\caption{Continued}
\end{figure*}

\begin{figure*}
\addtocounter{figure}{-1}
\plotone{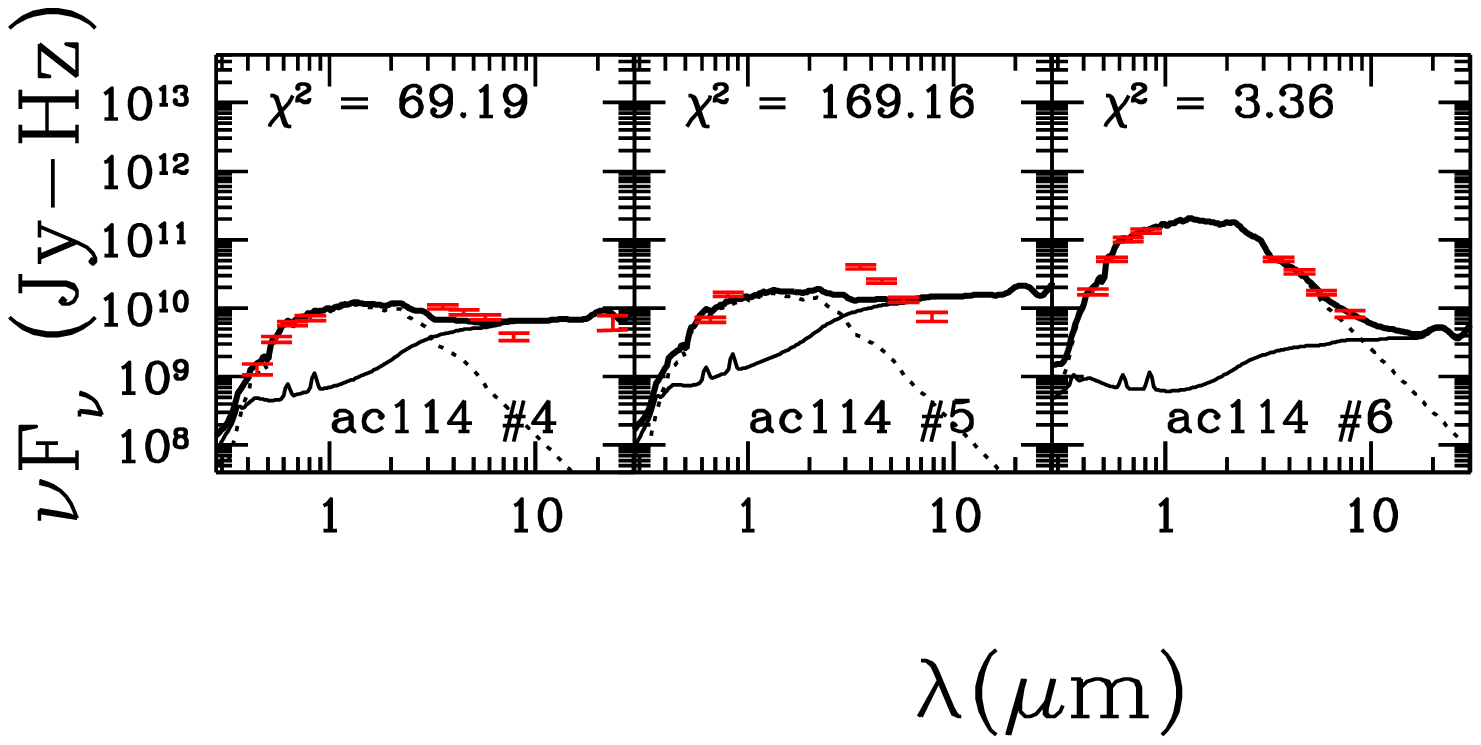}
\caption{Continued.  The poor fit to AC114\#5 indicates a bad
spectroscopic redshift.}
\end{figure*}

The model SEDs fit to 25 of the 488 spectroscopically-identified cluster
members are poorly matched to the measured photometry ($\chi^{2}>25$).
We determine photometric redshifts for all of the identified cluster
members, and in cases where the measured photometric redshifts are more than
$3\sigma$ away from the cluster redshift, we replace the spectroscopic
redshifts with photometric redshifts and repeat the fit.
In 11 cases, this procedure results in substantial improvements to the
fits ($\Delta\chi^{2}>12$, $\chi^{2}_{photo-z}<4$).
This suggests that some galaxies in the sample have erroneous spectroscopic
redshifts.  One such object is an X-ray source, identified as AC 114-5
by M06.  The redshift for this object was reported by
Couch et al.\ (2001; their galaxy \#365).  The spectra used by these
authors covered a relatively narrow wavelength range 
($8350{\rm \AA}<\lambda<8750{\rm \AA}$) and had moderately poor S/N.
We suspect that this combination of factors, in concert with a strong
prior in favor of cluster membership 
in the presence of a putative H$\alpha$ emission line
at the correct redshift, led \citet{couc01} to mis-identify the
[{\sc Oii}]$\lambda\lambda$4354 and [{\sc Oiii}]$\lambda\lambda$4363 emission
lines of a background quasar at $z=0.988$ as the [{\sc Nii}]$\lambda\lambda$6548
and H$\alpha$ emission lines, respectively, at the cluster redshift.
Four of the 5 objects flagged as having erroneous redshifts in AC 114 have
redshifts from \citet{couc01}.  Two of the four have redshifts from only
one emission line, and we have confirmed that both objects with redshifts
from multiple emission lines have plausible pairs of lines near the
photometric redshifts.  Furthermore, all of the objects with apparently
erroneous redshifts are quite faint, having $V\lesssim22$,
which makes acquiring high-S/N spectra difficult.
Our identification of objects with discrepant photometric
and spectroscopic redshifts as interlopers appears to be
reliable, and we eliminate the associated galaxies from further consideration.
The absence of AC 114-5 from the X-ray AGN sample has important repercussions,
which we discuss in Section \ref{secResults}.

\begin{figure*}
\plotone{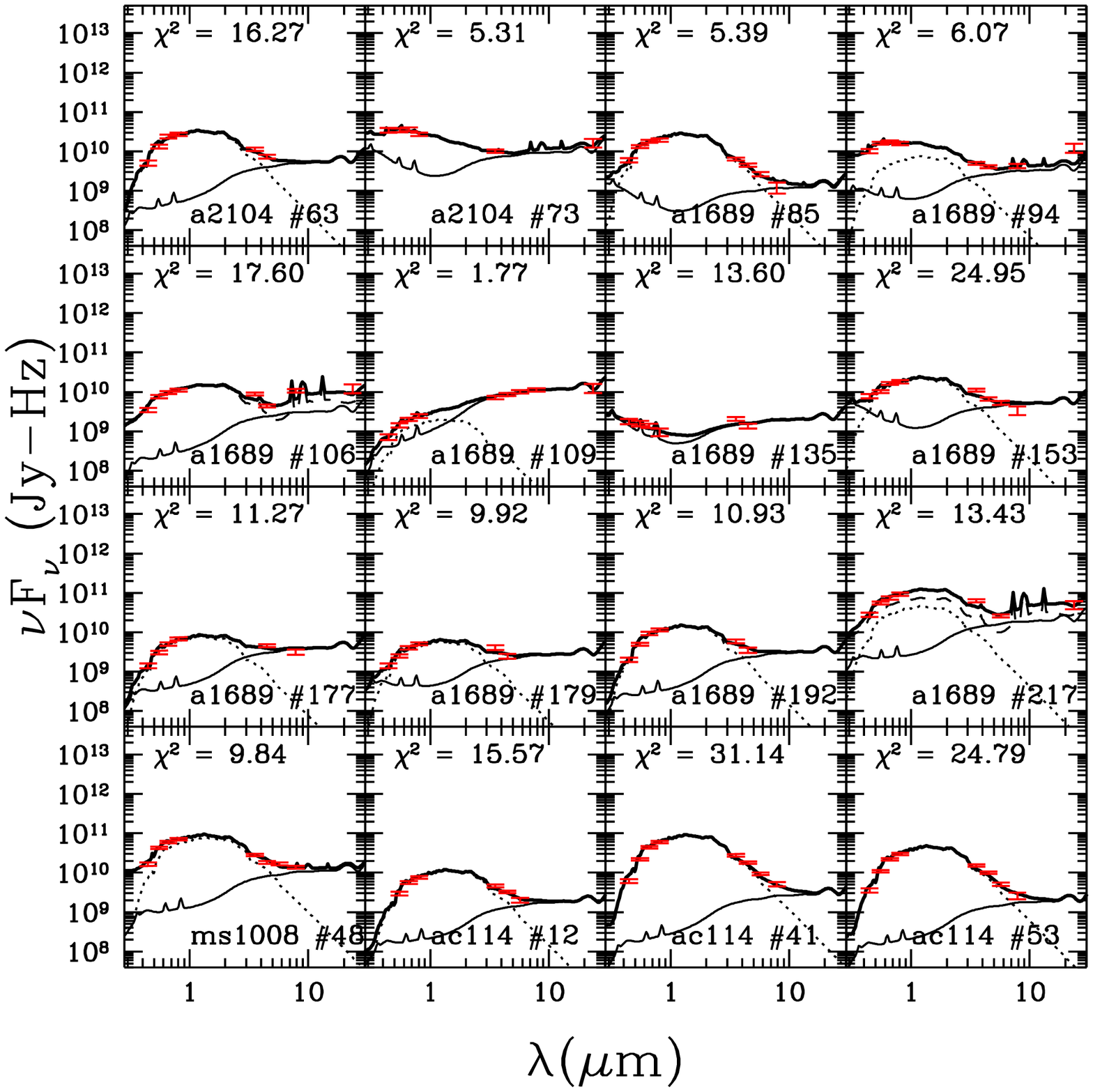}
\caption{Model SEDs for objects identified as IR AGNs which are not
also identified as X-ray AGNs.  Line types and bandpasses shown are the
same as in Figure \ref{figXrayFits}.
The object names indicated on each panel correspond to those
in Table \ref{tabAgn}.  See Section \ref{secModelSeds} for further details.
\label{figIrFits}}
\end{figure*}

\begin{figure*}
\plotone{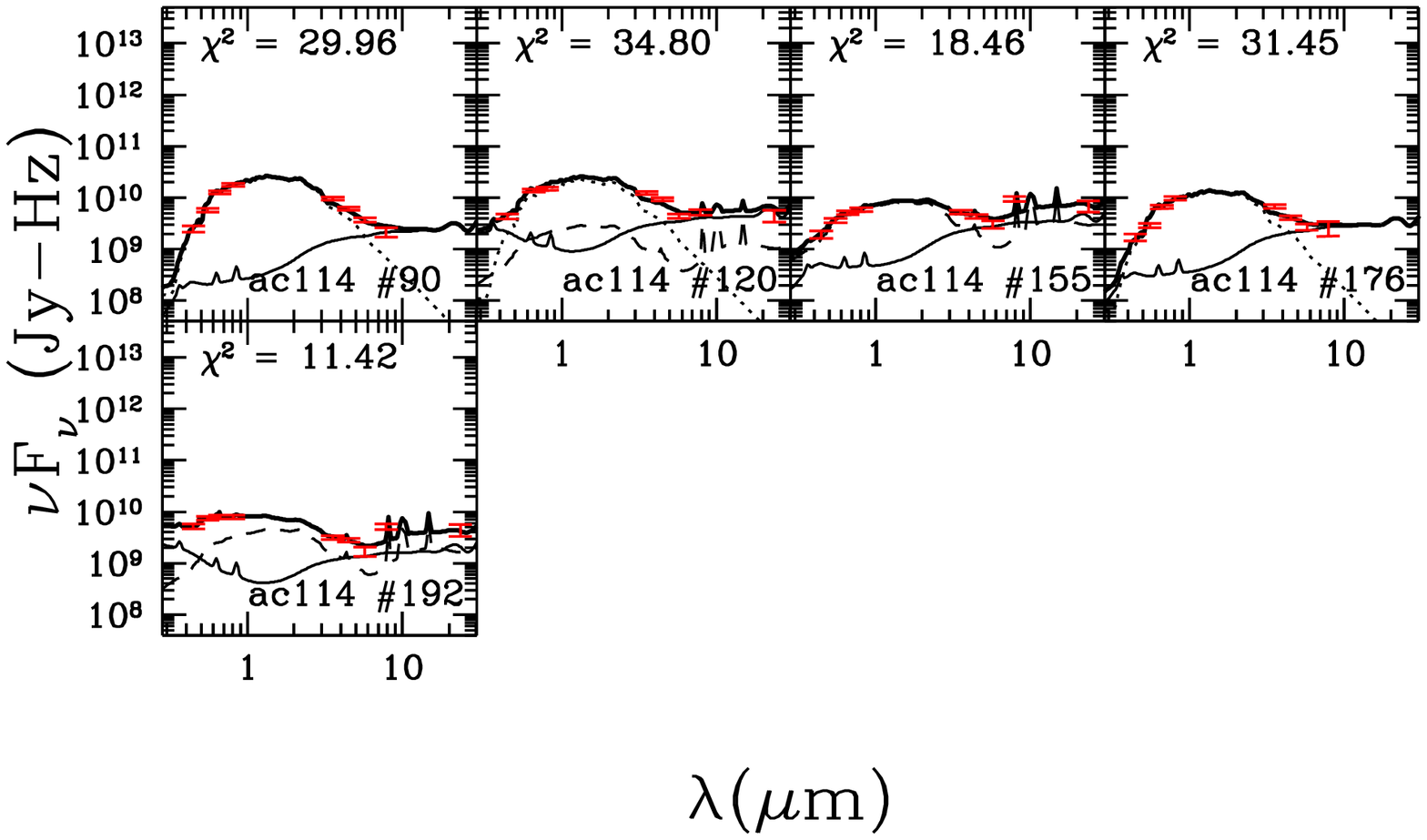}
\addtocounter{figure}{-1}
\caption{Continued}
\end{figure*}

\subsection{AGN Identification}\label{secAgn}
We consider AGNs selected
based on their X-ray luminosities, the shapes of their SEDs,
or both.  X-ray sources with 
$L_{\rm X}>10^{42}\ {\rm erg\ s^{-1}}$ are unambiguously AGNs,
but a number of processes can 
produce X-ray luminosities in the $10^{40}$--
$10^{42}\ {\rm erg\ s^{-1}}$ range.
These include LMXBs, HMXBs and a galaxy's extended, diffuse halo gas.
The integrated X-ray luminosities of LMXBs and hot halo both
correlate strongly with stellar mass, as measured by the
galaxy's $K$-band luminosity \citep{kim04,sun07}, and
the luminosity from HMXBs correlates with SFR \citep{grim03}.
These correlations allow us to predict the X-ray luminosity of a normal
galaxy using only parameters that can be measured from the model SEDs.
Similar analyses were performed by \citet{siva08} and \citet{arno09},
who used $K$-band luminosities measured from 2MASS photometry rather
than luminosities estimated from model SEDs.

We measure $K$-band magnitudes from the model SEDs and determine
SFRs from the K-corrected $8\mu m$ and $24\mu m$ luminosities
of X-ray sources in each cluster.  We use $L_{K}$ and SFR in 
Eqns.\ \ref{eqLmxb}, \ref{eqHmxb} and \ref{eqCorona} to
predict the expected X-ray luminosities from the host galaxies of X-ray 
point sources identified by M06
(\citealt{kim04,grim03,sun07}, respectively).
The predictions for X-ray emission from a given galaxy due to
LMXBs, HMXBs and the thermal
halo are good to within $\sim0.3$ dex and are given by,
\begin{align}\label{eqLmxb}
&L_{\rm X}(LMXB; 0.3-8\ {\rm keV})= \nonumber \\
&[(0.20\pm0.08)\times10^{30}\ {\rm erg\ s^{-1}}] \frac{L_{K}}{L_{K,\odot}}
\end{align}
\begin{equation}\label{eqHmxb}
L_{\rm X}(HMXB)=2.6\times10^{39}\ {\rm erg\ s^{-1}} \biggl[\frac{SFR}{M_{\odot}\ yr^{-1}}\biggr]^{1.7}
\end{equation}
\begin{align}\label{eqCorona}
&L_{\rm X}(thermal; 0.5-2\ {\rm keV})= \nonumber \\
&2.5\times10^{39}\ {\rm erg\ s^{-1}} \biggl[\frac{L_{K_{s}}}{10^{11}L_{\odot}}\biggr]^{1.63\pm0.13}
\end{align}
where $L_{K}$ and $L_{K_{s}}$ are the galaxy's luminosities in the $K$- and
$K_{s}$-filters.  Each relation is given in slightly different 
energy ranges, none of which coincide with the range used by
M06.  This problem is especially severe for Eqn.\ \ref{eqHmxb},
because \citet{grim03} take their X-ray fluxes from various sources 
in the literature without converting them to a common energy range.
They claim that the resulting uncertainty is
small because the scatter in the relation is much larger
than the bandpass corrections.  Fortunately, even if this were not the case,
the HMXB contribution to the total predicted X-ray luminosities
is small for the SFRs typical of cluster galaxies
($<10\ M_{\odot}\ {\rm yr^{-1}}$).  The contribution from thermal emission to
the soft X-ray luminosity can be significant,
dominating the LMXB component for
$L_{soft}\gtrsim6\times10^{40}\ {\rm erg\ s^{-1}}$.  This 
transition luminosity depends
on the specific form adopted in Eqn.\ \ref{eqCorona}.  \citet{mulc10}
found that $L_{\rm X}(corona)\propto L_{K}^{3.9\pm0.4}$ for field galaxies,
which differs significantly from the results of \citet{sun07}.
While the \citet{mulc10} relation is not strictly applicable to our sample, 
the difference between cluster and field galaxies suggests
that the thermal X-ray emission from a galaxy's halo depends on its
environment.  Such a variation introduces a systematic uncertainty
in $L_{\rm X}(corona)$ of up to 0.8 dex at $L_{K}=4\times10^{11}L_{\odot}$.
Hereafter we neglect this uncertainty, as its effect in a
given cluster is impossible to quantify given the data presently available.

\begin{figure}
\epsscale{1.0}
\plotone{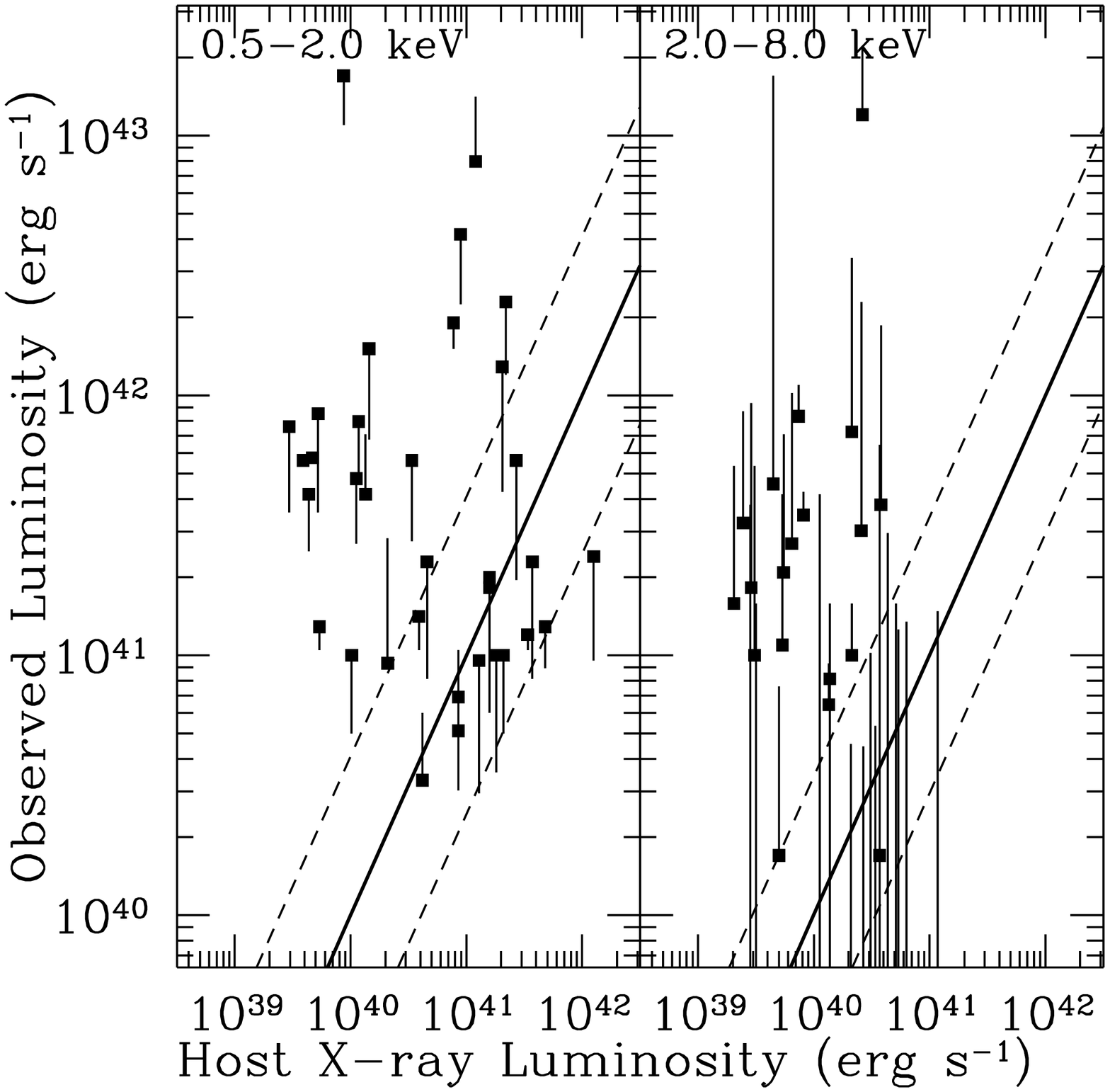}
\caption{Comparison of the X-ray luminosities of X-ray point
sources from M06 (y-axis) to the predicted X-ray 
luminosities of their host galaxies (x-axis).  Points show
the measured luminosities, and the ``tails''
connect each source to the luminosity estimated by separating its
$0.5-8.0\ {\rm keV}$  X-ray luminosity into soft and hard components
using a $\Gamma=1.7$ power-law.
The length of the tail indicates how well the measured photon energies
are described by a $\Gamma=1.7$ power law, and consequently describes
the systematic uncertainty on the quoted $L_{\rm X}$.
Long tails belong to objects poorly described by a $\Gamma=1.7$ 
power law.  {\it Heavy lines}
mark the line of equality ($L_{\rm X}=L_{host}$), 
and the {\it dashed lines} show the 
$\pm0.7$~dex scatter about the empirical relations used to
predict the X-ray luminosity of a given host galaxy.
See Section \ref{secAgn} for the method used to predict X-ray luminosities
of normal galaxies.
\label{figXrayLum}}
\end{figure}

We convert Eqns.\ \ref{eqLmxb}-\ref{eqCorona} to determine luminosities in
the soft X-ray (0.5-2 keV)
and hard X-ray (2-8 keV) bands, assuming a $\Gamma=1.7$ power law
for the LMXB and HMXB relations.  We further assume that the \citet{grim03}
relation corresponds to luminosities in the 2-10 keV range and
that the thermal emission from the $kT=0.7\ {\rm keV}$ halo gas
is negligible in the hard X-ray band.
The X-ray luminosities reported by
M06 and our estimates of the systematic uncertainties in these
luminosities associated with the choice of energy correction factor
(ECF) are shown in Figure \ref{figXrayLum}, along with 
the predicted luminosities from the host galaxies.
Many of the reported point sources require an AGN component, but several
of the M06 point sources have very massive host
galaxies, and their observed fluxes may arise entirely from
non-AGN sources.  

M06 selected 40 X-ray point sources with reliable detections above
the extended emission from the surrounding ICM ($N_{count}\geq 5$).
Of these 40 sources, they identify 35 as probable AGNs.
We have sufficient photometry to construct reliable model SEDs for 35 M06
X-ray point sources.
The remaining 5 M06 point sources either
lack enough data to produce a reliable model SED or fall outside
the $R$-band field of view.
We find that 23 of these 35 sources have X-ray luminosities more
than $1\sigma$ greater than the predicted host luminosity.  Henceforth,
we will call these objects X-ray AGNs.
The systematic flux error estimates in
Figure \ref{figXrayLum} indicate that many X-ray AGNs have photon
energy distributions that are poorly matched to the $\Gamma=1.7$ power-law
assumed by M06.
Three such AGNs are close to the boundary
separating probable AGNs from more ambiguous cases and have too large
a soft X-ray flux compared to their hard X-ray flux to be consistent with
a $\Gamma=1.7$ power law.  M06 did not correct for X-ray
absorption, and in the cases where the ratio of soft to hard X-ray photons
is too low for a $\Gamma=1.7$
power law, absorption may explain the apparent discrepancy.
However, objects whose soft X-ray fluxes are unexpectedly large compared to
the total cannot result from absorption.

Many narrow-line Seyfert 1 galaxies (NLS1) show excess soft X-ray emission
\citep{arna85}.  However, only one X-ray source identified by M06
is a NLS1 (their Abell 644 \#1),
so the soft X-ray excess common to NLS1s cannot explain
the presence of excess soft X-ray emission in 13 X-ray sources with AGN-like
luminosities.
Alternative explanations include soft X-rays arising
from gas that is photoionized by an obscured AGN (e.g.\
\citealt{ghos07}), poor signal-to-noise in the X-ray, and thermal
emission from hot gas.  The
ECF used to convert soft X-ray photons to incident
fluxes for $kT=0.7\ {\rm keV}$ thermal bremsstrahlung 
(assumed by \citealt{sun07}) is larger than the
ECF for a $\Gamma=1.7$ power law by approximately 10\%.  This implies
that two of the
three suspect X-ray AGNs have luminosities sufficiently close to the threshold
that they may reasonably be mis-classified galaxies.  This yields a possible
contamination in the X-ray AGN sample of approximately 10\%,
which is comparable to the estimated contamination of the IR AGN sample
(see below).

In comparison to our sample of 23 X-ray AGNs from a parent sample of
35 X-ray point sources with complete photometry,
M06 found that 35 of their
40 point sources had X-ray luminosities consistent with AGNs.  The
larger fraction of AGNs reported by M06
may be attributed to their use of
$L_{\rm X}$--$L_{B}$ relations, which show larger scatter than the $K$-band
relations.  We also introduce some uncertainty by estimating $L_{K}$
from the model SEDs, but this uncertainty is small
($\sim10\%$) compared to the scatter in the $L_{\rm X}$--$L_{K}$ relation.
An additional difference is that M06 considered
the two luminosity components
separately and did not compare their sum to the measured luminosities,
This was done subsequently by \citet{siva08} and \citet{arno09}
in their studies of AGNs in low-redshift groups
and clusters of galaxies.  Their analyses are much closer to our
method, and their samples 
included some of the clusters in our sample (Abell 3128, 3125 and 644).

An alternative method to identify AGNs is to use the 
distinctive shape of their SEDs, particularly
in the MIR (e.g.\ \citealt{marc04,ster05,rich06}; A10).
This approach can identify
AGNs behind gas column densities large enough to obscure 
even the X-rays emitted by an AGN.  Such an AGN sample has very different
selection criteria and biases than an X-ray selected sample,
and combining the two results in more
complete AGN identification.

We identify AGNs from their SEDs
by comparing the goodness-of-fit of two sets of
model templates.  The first set uses only
the normal galaxy templates.  The other also includes the AGN template.
We determine whether a given galaxy requires an AGN component in its model
SED by applying a threshold on the likelihood ratio, $\rho$,
\begin{equation}\label{eqLikelihood}
\rho=\frac{\exp[-\chi^{2}(gal)/2]}{\exp[-\chi^{2}(gal+AGN)/2]}
\end{equation}
where $\chi^{2}(gal)$ and $\chi^{2}(gal+AGN)$
are goodnesses-of-fit for a model with only the A10 galaxy templates and
for a model that includes an additional AGN component, respectively.
AGNs are those objects whose $\rho$ is smaller than
a pre-determined selection limit, $\rho_{max}$,
established by Monte Carlo simulations of normal galaxies.

\begin{figure}
\epsscale{1.0}
\plotone{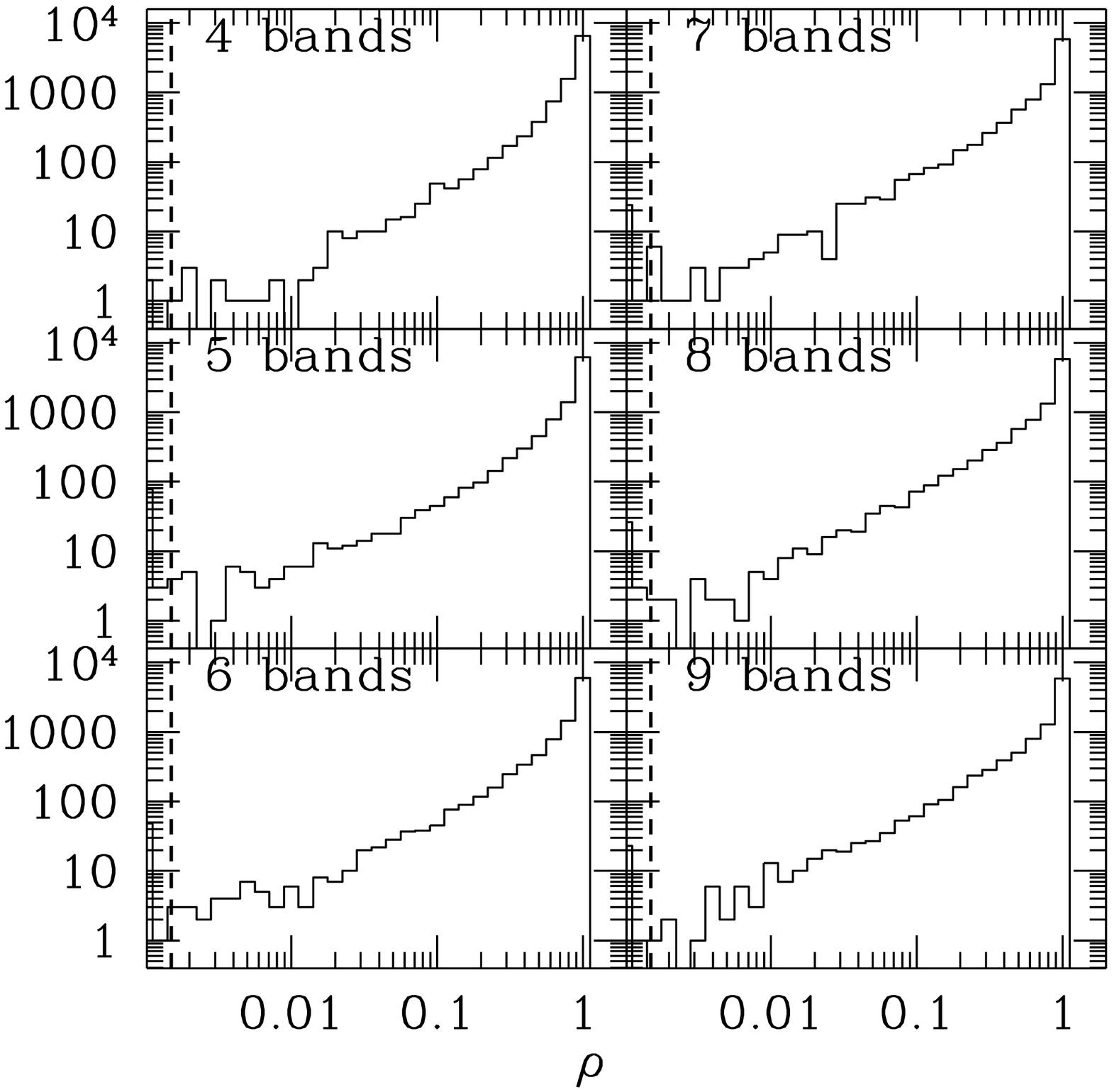}
\caption{Likelihood ratio ($\rho$) distributions for fits to
model galaxies with no AGN component after photometric errors 
have been added.
Each panel shows the distributions resulting from examining 10,000
normal galaxies.
Galaxies with $\rho=1$ are well-fit by the three normal galaxy templates and
do not require an AGN component.  An object with $\rho=0$
would be perfectly fit by the 4-template model SED
and have $\chi^{2}_{gal+AGN}=0$.
The dashed line indicates the selection
threshold, $\rho_{max}$ used to identify IR AGNs.
See Section \ref{secAgn} for further details.
\label{figLikelihood}}
\end{figure}

We created artificial galaxy photometry to determine an
appropriate $\rho_{max}$ by combining the three 
galaxy templates of A10 in proportions that reflect the
template luminosity distributions in real cluster members.
We introduced Gaussian photometric errors comparable to the photometric
uncertainties in our real data (0.07 mag) to the fluxes given
by the model SEDs.  We also allowed occasional
catastrophic errors of up to 0.3 dex.
The artificial galaxy photometry did not include upper
limits, which we also neglected when constructing model SEDs of real galaxies.
We fit the artificial galaxies with
two models.  The first model excluded the AGN component from the fit,
while the second component included it.  
The likelihood ratio distributions computed from the goodness-of-fit
results for the two different models are shown
in Figure \ref{figLikelihood}.  These distributions show the
probability that a pure galaxy will be erroneously classified as an
AGN due to the presence of photometric errors.
The similarity of the different distributions, 
even based on only 4 photometric bands, indicates 
that a single $\rho_{max}$ can be used to select AGNs from
among all galaxies in our sample.

\begin{figure}
\epsscale{1.0}
\plotone{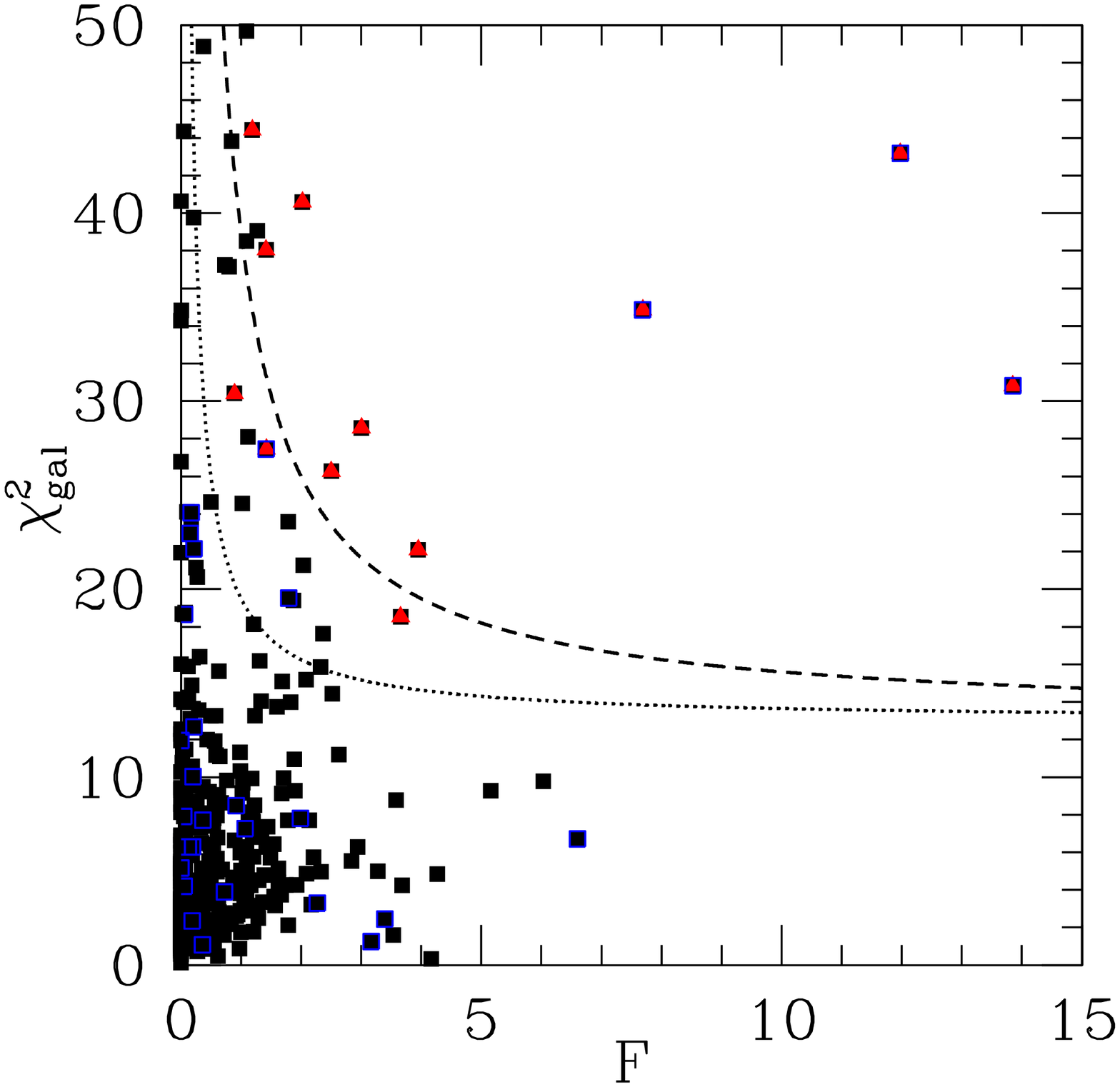}
\caption{Application of F-statistics for AGN identification based on fits
of galaxy-only and galaxy-plus-AGN model SEDs to measured photometry.
{\it Red triangles} show AGNs identified using the $\rho$ threshold
shown in Figure \ref{figLikelihood} (IR AGNs).
{\it Open blue squares} show X-ray AGNs, and 
{\it solid black squares} show ``normal'' cluster members.  The {\it dotted}
and {\it dashed} curves show the $\rho$ thresholds for objects having N=6
and N=9 flux measurements, respectively.  Objects above their corresponding
selection boundaries are identified as AGNs, provided that they pass a
$\chi^{2}_{\nu}$ cut.
\label{figFstat}}
\end{figure}

We also identify AGNs based on the F-statistics of
the two model SED fits described above.
Figure \ref{figFstat} shows the F-statistic as a function of
$\chi^{2}(gal)$ for X-ray AGNs selected using Figure \ref{figXrayLum},
AGNs selected using likelihood ratios, and ``normal'' cluster
members.  The F-statistic is given by,
\begin{equation}\label{eqFstat}
F=\frac{\Delta\chi^{2}/2}{\chi^{2}_{\nu}(gal+AGN)}
\end{equation}
where $\Delta\chi^{2}$ is the (absolute) change in the total $\chi^{2}$ after
introducing the AGN component to the fit.  In addition to the
galaxies that are well-fit by the galaxy-only model and not
substantially improved by the addition of an AGN component, there are
objects with large $\chi^{2}(gal)$
but small $F$, and objects with large $F$ but small $\chi^{2}(gal)$.
Neither of the latter two categories contain objects likely to be AGNs from
the point-of-view of the model SEDs.
The most luminous X-ray AGNs have both large F and large
$\chi^{2}(gal)$.  These are clearly identified as AGNs by the model SEDs,
and less luminous X-ray AGNs can be
found with increasing density toward the normal galaxy locus at the
origin of Figure \ref{figFstat}.
The dotted and dashed lines in the Figure correspond to
the $\rho<\rho_{max}$ selection boundaries for N=6 and N=9 flux
measurements, respectively.  Some objects above the N=9
line are not selected as IR AGNs because they fail a 
cut on the overall goodness-of-fit, which requires
$\chi^{2}_{\nu}(gal+AGN)<5$.  We could define an AGN selection region in Figure
\ref{figFstat}, but due to the
non-uniformity of our photometric data, this would result in different
effective cuts in $\Delta\chi^{2}$ between different clusters and
between objects in individual clusters.  Furthermore, we find that only 3
AGNs identified using likelihood ratios fall into the suspect part of
Figure 4 with $F\approx1$.  This level of contamination ($\sim10\%$)
is consistent with the estimated purity of the X-ray AGNs,
which we deem to be acceptable.  Therefore, for the rest of
this work, we rely on the more simplistic likelihood ratio threshold
to identify AGNs.

\begin{figure*}
\epsscale{1.0}
\plotone{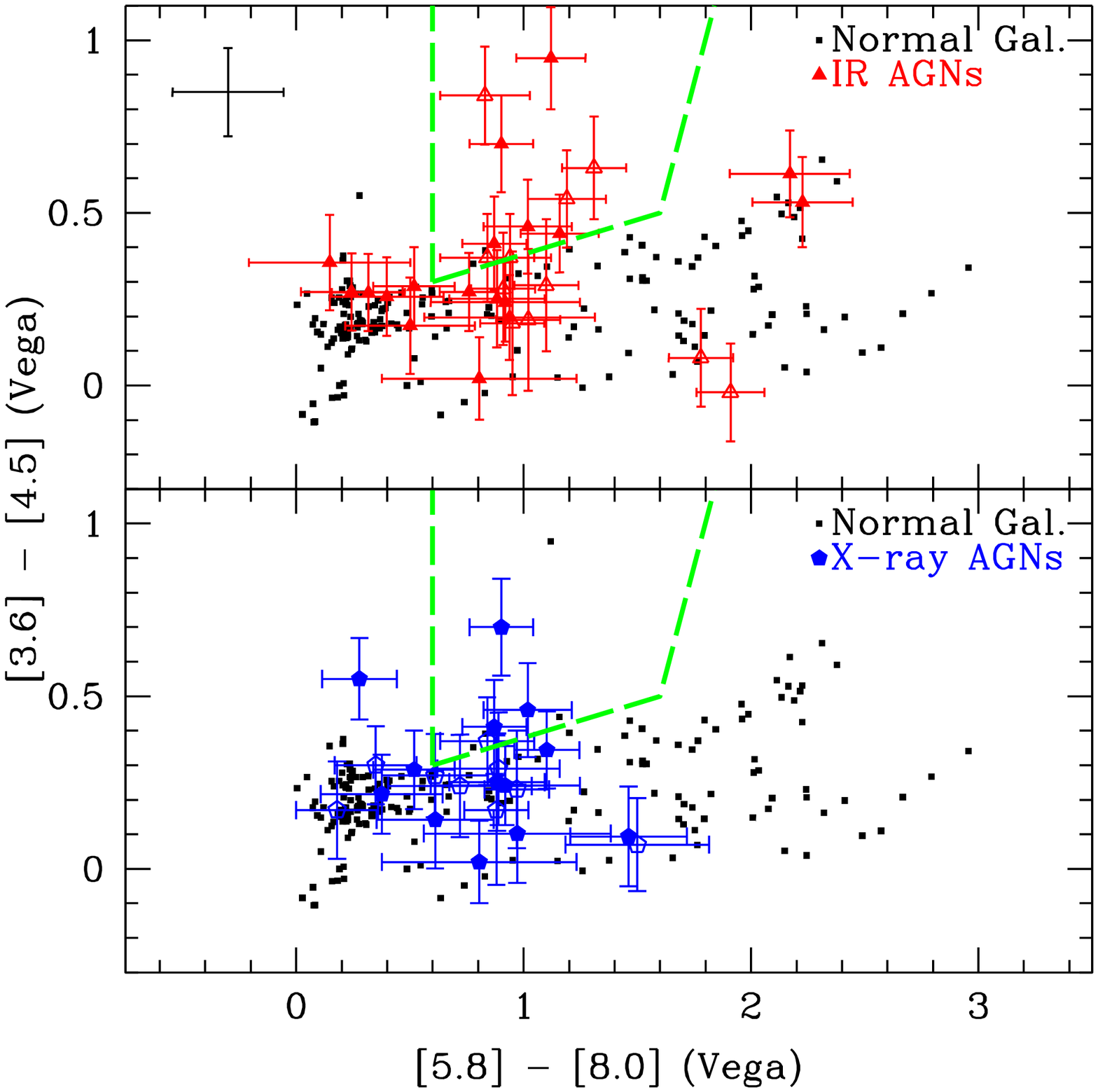}
\caption{Positions of both IR AGNs (red triangles, upper panel) and 
X-ray AGNs (blue pentagons, lower panel) on the \protect\citet{ster05} 
AGN selection diagram.  The dashed, green trapezoid marks the Stern wedge.
Filled symbols mark AGNs whose colors were determined
using only measured magnitudes; open symbols show colors determined
using model SEDs to estimate magnitudes in missing IRAC bands.
No model colors were constructed for normal galaxies.
Passive galaxies are located in the lower-left corner, normal
star-forming galaxies are found in the center, and (U)LIRGs occupy the
upper-right part of the galaxy sequence.  The error bars in the
upper-left show the color uncertainty for a typical cluster member,
including AGNs.
\label{figStern}}
\end{figure*}

Likelihood ratio selection of AGNs using SED fitting
is most sensitive to the shape of the MIR SED, so we refer
to AGNs so identified as IR AGNs.
We find 29 IR AGNs using a selection boundary at the 
99.8\% confidence interval of the merged 
$\rho$ distribution ($\rho_{max}=1.5\times10^{-3}$).
Table \ref{tabAgn} lists both X-ray and IR AGNs, their
luminosities, and the basic parameters of their host
galaxies.
IR AGN selection recovers 5 of 7 AGNs (71\%) identified via the
Stern wedge (\citealt{ster05}, see Figure \ref{figStern}) and 8 of
the 23 X-ray AGNs.  The galaxies in the Stern wedge that are not
selected from their SED fits
fall just inside the boundary of the wedge, so 
they may be normal galaxies shifted into the wedge by photometric errors.
\citet{gorj08} find that 35\% of X-ray sources in the Bo\"otes field
of the NOAO Deep Wide Field Survey 
($f_{\rm X}>8\times10^{-15}\ {\rm erg\ s^{-1}\ cm^{-2}}$) with detections in
all 4 IRAC bands fall outside the Stern wedge, and Figure 14 of A10
shows that a substantial fraction of the point-source (luminous)
AGNs in their sample fall outside the wedge as well.  Given the high
luminosities in both of these samples, it is perhaps not surprising that
most of the lower-luminosity AGNs common in galaxy clusters
fall outside the Stern wedge.
For $\rho_{max}=1.5\times10^{-3}$
and the size of our sample (488 galaxies),
we expect on average one false-positive AGN identification
and 3 or fewer false-positives at 98\% confidence,
implying $>90\%$ purity in our IR AGN sample.

We estimate the completeness of the IR AGN sample as a function of
the reddening of the AGN template and luminosity using Monte Carlo
simulations.
We construct model AGN SEDs by injecting an AGN component with some
luminosity and reddening into artificial galaxy photometry, which we
generate using the Monte Carlo techniques described above.
We estimate the completeness of the IR AGN sample from the fraction of 
such AGNs recovered.  The completeness depends
strongly on the luminosity of the AGN component.  We only reliably
identify AGNs with $L_{bol}\gtrsim 7\times 10^{10}L_{\odot}$.
The completeness depends only weakly on $E(B-V)$.  There are
measurable differences only for AGNs with
$E(B-V)>2$.  For our observed wavelengths, AGN identification depends
 most strongly on the shape of the MIR SED, which is insensitive to modest
amounts of reddening.  The full
dependence of completeness on $L_{bol}$ and $E(B-V)$ is listed in Table
\ref{tabComplete}.

We caution that both our AGN identification and the analysis
below were conducted using the fixed AGN template derived by A10.
While this template is dominated by luminous AGNs, AGNs of
all luminosities were used in its construction, and in
some sense it represents the optimal median AGN SED.
There is some evidence that
AGNs with low Eddington ratios ($L_{bol}/L_{Edd}$) are systematically
weaker in the UV and the MIR than higher $L_{bol}/L_{Edd}$ AGNs.
This appears to become important at $L_{bol}/L_{Edd}\approx10^{-3}$
\citep{ho08}.  However, the UV weakness of such objects remains
a subject of debate (e.g.\ \citealt{ho99,ho08,dudi09,erac10}),
and the SEDs of AGNs appear to
all be quite similar out to $\lambda\approx20\mu m$, 
even in AGNs with accretion rates as low as $L_{bol}/L_{Edd}\approx10^{-3}$
(\citealt{ho08}, Figure 7).  Furthermore, the variable reddening
of the AGN component allowed by the models can account
for differing UV/visible flux ratios,
making the AGN component of the model SEDs flexible enough to mimic
AGNs with a wide variety of Eddington ratios.

Intrinsic variations in the AGN SED are one possible cause of
the absence of an important AGN component in the SEDs of many
X-ray AGNs, despite their
similar distributions in $L_{bol}$ (Section \ref{secResults}).
Another possible explanation is that the nuclear MIR emission
from many X-ray AGNs is overwhelmed by star-formation in their
host galaxies.  We find that X-ray AGNs with
$L_{\rm X}>10^{42}\ {\rm erg\ s^{-1}}$ that are also identified as
IR AGNs have no measurable star-formation, while those not identified
in the IR have $\langle SFR \rangle = 0.3\ M_{\odot}\ {\rm yr^{-1}}$.
This may be a selection effect, since nuclear MIR emission
is not subtracted before computing SFRs in
galaxies not identified as IR AGNs.  However, it appears that the
balance between SFR and nuclear emission is an important factor in
determining whether a given X-ray source will be identified as an IR AGN.

Also of concern is the MIR emission exhibited by some normal galaxies
which is clearly not associated 
with star-formation (e.g.\ \citealt{verl09,kels10}).
The strength of the diffuse interstellar dust
emission relative to star formation varies from galaxy to galaxy depending
on the populations of AGB stars, which can produce and heat dust \citep{kels10},
and field B-stars (including HB stars),
which produce UV light that can both heat dust grains
and excite PAH emission in the diffuse ISM (e.g.\ \citealt{li02}).  These
effects could mimic the presence of an AGN, particularly in
passively-evolving galaxies, which the A10 templates predict should
decline strictly as a $\nu F_{\nu}\propto\nu^{2.5}$ power-law.
Given the limited data available to constrain MIR emission not associated
with either an AGN or a star-forming region and the
as-yet uncertain magnitude of the associated variations, we neglect any
potential effects on our AGN identification.  However, potential
sources of MIR
emission not accounted for by the A10 templates, especially emission from
dust heated by old stars in passive galaxies, remain a potentially important 
systematic uncertainty.

\subsection{Stellar Masses}\label{secMass}
Stellar population synthesis modeling provides a means to estimate
stellar masses in the absence of detailed spectra.  \citet{bell01}
construct model spectra of galaxies for a wide variety of
stellar masses, SFRs,
metallicities and stellar initial mass functions (IMFs)
to convert colors to mass-to-light ratios (M/L).
Their models assume a mass-dependent formation epoch with
bursty star-formation histories, which is appropriate for the
spiral galaxies they study.  Figure 9 of
\citet{bell01} makes it clear, however, that their results also
robustly estimate M/L 
for passively evolving galaxies.  In fact, the scatter about
the mean M/L tends to decrease for redder systems because the stochasticity
of the star-formation history becomes less 
important in galaxies that experienced
their last burst of star-formation in the distant past.

Bell \& de Jong provide a table of coefficients ($a_{\lambda}$,$b_{\lambda}$)
relating M/L for a galaxy to its color,
\begin{equation}\label{eqML}
\log_{10}\left(M/L_{\lambda}\right)=a_{\lambda}+b_{\lambda}\times{\rm color}
\end{equation}
where {\it color} is measured in the bands for which
$a_{\lambda}$ and $b_{\lambda}$ were determined.
We adopt the coefficients
appropriate for Solar metallicity computed with the
\citet{bruz03} population synthesis code and the scaled Salpeter IMF suggested
by \citet{bell01}, who report that a modified Salpeter IMF with total mass 
$M'=0.7M_{Salpeter}$ yields
the best agreement with the Tully-Fisher relation.  Once
we select an appropriate ($a_{\lambda}$,$b_{\lambda}$) pair, it is
straightforward to compute stellar masses from the visible photometry.
However, we must first subtract the AGN component of the model SED
in sources identified as IR AGNs before computing colors.

\begin{figure}
\epsscale{1.0}
\plotone{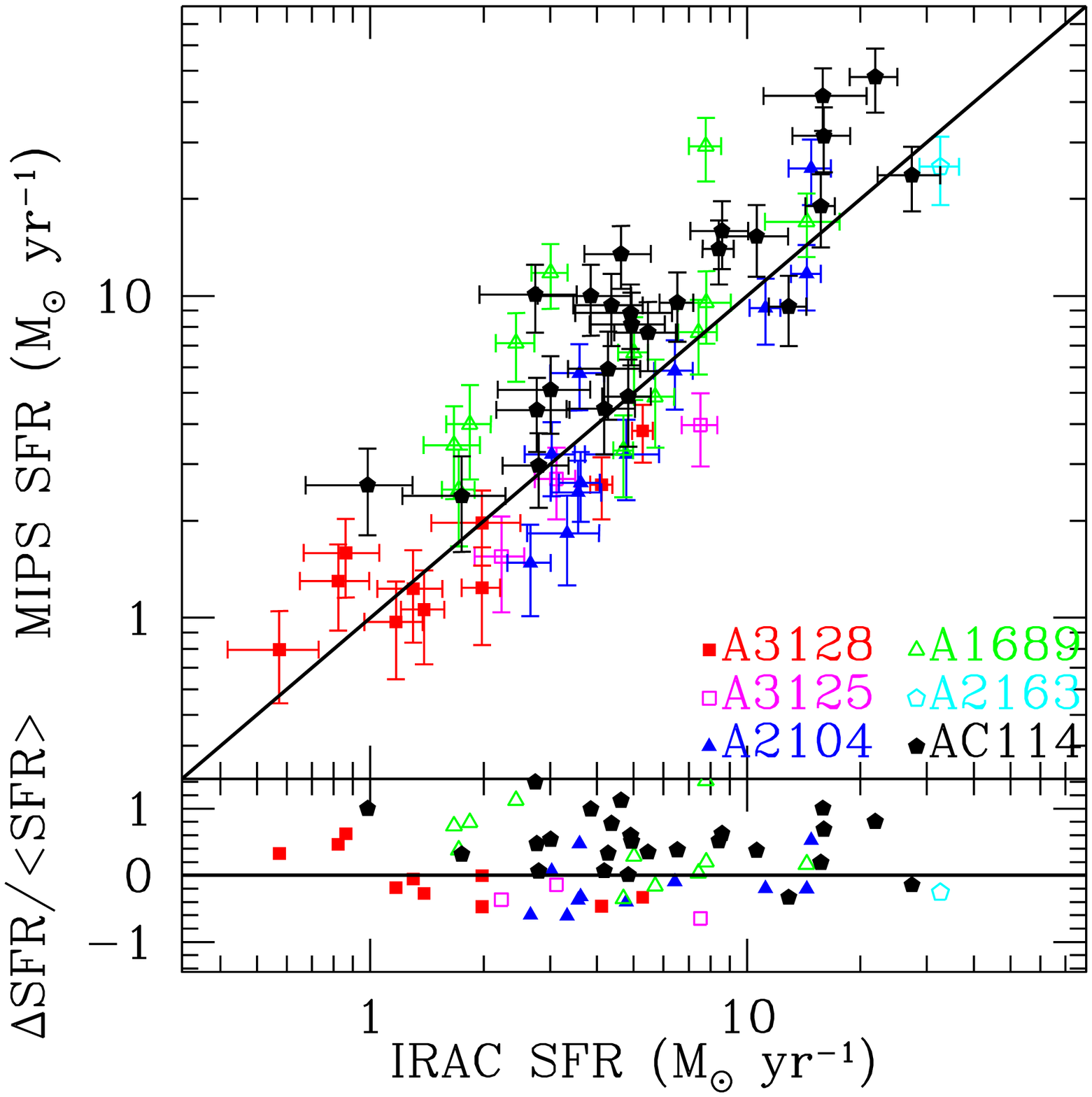}
\caption{Comparison of SFRs determined using the
IRAC and MIPS relations of \protect\citet{zhu08}.
The solid lines in both panels
denote equality.  The bottom panel shows the fractional
residuals between the two star-formation indicators.
The MIPS SFRs are biased high with respect to IRAC SFRs by
0.1 dex, which is significant at $2.4\sigma$ but smaller than
the intrinsic scatter (0.2 dex).
\label{figSFR}}
\end{figure}

The uncertainty introduced by the AGN subtraction
is a combination of the fractional uncertainty in the
contribution of the AGN template to the model SED, which is determined
by the fit, and the uncertainty in the AGN template itself.  To
measure the uncertainty in the template, we examined
1644 luminous quasars with spectroscopic redshifts from the AGN and Galaxy
Evolution Survey (AGES; \citealt{kochIP}) and determined the variation
in their measured photometry about their best-fit model SEDs.
Using these measurements, we constructed an RMS SED
for AGNs and averaged it across each of
the bandpasses we employ.  The uncertainty in the AGN
correction resulting from intrinsic variation about the AGN
template is 10\% except at $24\mu m$,
where there are too few $z=0$ quasars to make a meaningful comparison.
The uncertainty in the AGN correction at $24\mu m$ is therefore large,
but it can be 
constrained by the relatively good agreement of the $8\mu m$ and $24\mu m$
SFRs (Figure \ref{figSFR}).

In galaxies with no genuine nuclear activity the AGN template can
correct for variations in stellar populations relative to the
templates, intrinsic extinction, or
errors in the measured photometry.  Subtraction of the AGN
component under these circumstances would result in under-estimated
stellar masses and SFRs, while failure to subtract the AGN component
in a genuine, low-luminosity AGN would cause the measured SFRs
of their host galaxies to be biased toward higher values.
However, the ambiguity between a genuine, low-luminosity AGN and an
apparent AGN component introduced to correct for
photometric errors (Section \ref{secAgn}) renders
any attempt to subtract the AGN component in such cases suspect.
Therefore, in normal galaxies and in X-ray AGNs not identified as IR AGNs,
no AGN correction is applied.
We accept the inherent bias to avoid introducing ambiguous AGN
corrections, which would be much more difficult to interpret.

The \citet{bell01} calibrations are reported for
rest-frame colors, so we need
K-corrections for each cluster member to convert the measured
magnitudes to the rest frame.
We calculate the K-corrections from the model SEDs
returned by the A10
fitting routines.  Uncertainties on K-corrections cannot be directly
determined from the uncertainties in the model components because
K-corrections depend non-linearly on these uncertainties.
Therefore, we recombine the components of each model SED
in proportion to the uncertainties in their contributions to the total
model flux.  This results in a series of temporary model SEDs.
We then calculate the K-corrections implied by these temporary model SEDs
and measure their dispersions to estimate the uncertainties
in the K-corrections returned by the original model SED.

The systematic uncertainty on the stellar masses calculated from
Eqn. \ref{eqML} can be estimated
by comparing the fiducial masses with masses derived using different
assumptions.
We estimate the typical systematic uncertainty
in stellar mass, listed in Table \ref{tabAgn},
to be 0.2 dex.  These uncertainties are derived by
measuring the difference between the fiducial masses and those
determined using coefficients appropriate for 
the {\sc{P\'egase}} population synthesis models with
a Salpeter IMF.  \citet{conr09} studied the ability
of different models to reproduce the observed colors of
stellar populations in globular clusters
and found that systematic uncertainties on
stellar masses derived from population synthesis codes typically
reach or exceed 0.3 dex.

\subsection{Star-Formation Rates}\label{secSFR}
We measure SFRs from our AGN-corrected MIR photometry using the empirical
relations of \citet{zhu08}, which have been
determined for both the IRAC $8\mu m$ and
the MIPS $24\mu m$ bands using the same calibration sample.  While
the contribution of the stellar continuum to the observed $24\mu m$
luminosity is negligible,
the Rayleigh-Jeans tail of the stellar continuum emission can make an
important contribution to the integrated flux at $8\mu m$, especially in
galaxies with the low SFRs typical
in clusters.  The method used to subtract this contribution is an
important systematic uncertainty in the SFR calculation.
\citet{zhu08} assume
that the contribution of the stellar continuum at $8\mu m$ can be described by
$L_{\nu}^{stellar}(8\mu m)=0.232 L_{\nu}(3.5\mu m)$, as derived from
the models of \citet{helo04}.  Under this assumption,
\citet{zhu08} derive luminosity--SFR relations appropriate for a Salpeter IMF,
\begin{equation}\label{eqIracSf}
SFR(M_{\odot}\ {\rm yr}^{-1})=\frac{\nu L_{\nu}^{dust}(8\mu m)}{1.58\times10^{9}L_{\odot}}
\end{equation}
\begin{equation}\label{eqMipsSf}
SFR(M_{\odot}\ {\rm yr}^{-1})=\frac{\nu L_{\nu}(24\mu m)}{7.15\times10^{8}L_{\odot}}
\end{equation}
where $L_{\nu}^{dust}(8\mu m)$ is determined by subtracting
$L_{\nu}^{stellar}(8\mu m)$ from the the measured $8\mu m$ luminosity.
Sim\~oes-Lopes et al.\ ({\it in preparation}) find that
$L_{\nu}^{stellar}(8\mu m)=0.269 L_{\nu}(3.5\mu m)$ provides a better
estimate for their sample of nearby, early-type galaxies with no
dust and conclude that the difference in their result compared
to \citet{helo04} is due to metallicity.
Another important systematic uncertainty in SFRs derived from PAHs
is the dependence of the PAH abundance on metallicity \citep{calz07}
because lower
metallicity systems have fewer PAHs and therefore weaker $8\mu m$
emission at fixed SFR.  This second effect is negligible for the 
high-mass---and therefore metal-rich---galaxies we consider.  We neglect both
metallicity- and mass-dependent effects for the remainder of our
analysis.  Instead, we follow \citet{zhu08} and assume that
$L_{\nu}^{stellar}(8\mu m)=0.232 L_{\nu}(3.5\mu m)$.  We derive
SFRs from Eqns.\ \ref{eqIracSf} and \ref{eqMipsSf}.
For galaxies having measurable ($>3\sigma$) SFRs from both IRAC and MIPS,
we take a geometric mean of the two; otherwise, we use whichever
SFR measurement is available.  The resulting SFRs for AGNs are
summarized in Table \ref{tabAgn}.

Equations \ref{eqIracSf} and
\ref{eqMipsSf} were derived using the extinction-corrected H$\alpha$
luminosity of the associated galaxies.  The MIPS SFR determined
from Eqn.\ \ref{eqMipsSf} for a galaxy with 
$\nu L_{\nu}=7.15\times10^{9} L_{\odot}$ is $\approx0.6$ dex 
larger than the SFR derived from the \citet{calz07}
relation, which was calibrated using the Pa$\alpha$ emission line.  
\citet{calz07} used the Starburst99 IMF, and after accounting for this
difference, the resulting discrepancy is reduced to 0.4 dex.
The choice of SFR calibration therefore represents an important systematic
uncertainty in the measured SFRs.  The total systematic uncertainty in SFR
is indicated by the significant scatter (0.2 dex) and the small but
marginally significant offset (0.1 dex)
between the IRAC and MIPS SFRs in Figure \ref{figSFR}.
Since the offset is smaller than both the scatter about the line
of equality and the systematic uncertainty when comparing to the 
\citet{calz07} result, we neglect it below.
However, we caution that there remains 
a $\sim15\%$ uncertainty in our results associated with the discrepancy
between the IRAC and MIPS SFR indicators.

\section{Results}\label{secResults}
We identify 29 IR AGNs with likelihood ratios $\rho<\rho_{max}$.
We also confirm the presence of AGNs in 23 X-ray
point sources whose X-ray luminosities significantly exceed the
luminosities expected from their host galaxies.  Surprisingly, the
X-ray and IR AGN samples are largely disjoint: only 8 AGNs appear in
both.  Only the more
luminous IR AGNs appear in the X-ray AGN sample and
vice-versa.  While it is not surprising for faint X-ray AGNs
to drop out of the IR AGN sample, the absence of 
X-ray emission associated with many IR AGNs,
which require a moderately luminous AGN for a reliable
detection, is unexpected.
This may indicate either different selection biases in the
two methods or genuine, physical differences
between the AGNs selected by these techniques.

\subsection{Bolometric AGN Luminosities}\label{secBol}
In order to conduct a meaningful comparison of X-ray and IR AGNs,
we need to place them on a common luminosity
system.  The most obvious choice is the bolometric
AGN luminosity ($L_{bol}$), which also allows us to examine black
hole growth rates.

The A10 AGN template provides a natural means of
determining the bolometric luminosity ($L_{bol}$)
for IR AGNs, but the MIR luminosity in the
template comes from reprocessed dust emission, which would result in
double-counting the UV emission from the disk for AGNs viewed face-on
(\citealt{marc04}, hereafter M04; \citealt{rich06}).
We instead determine
$L_{bol}$ using a piecewise combination of the AGN model SED
and three power-laws.  We integrate the unreddened
A10 AGN template from Ly$\alpha$ to $1\mu m$, shortward of which the
template becomes uncertain due to absorption by 
the Ly$\alpha$ forest.  We estimate
the X-ray luminosity by integrating a $\Gamma=1.7$ power law
from 1--10 keV.  We estimate the extreme ultraviolet (EUV) luminosity by
integrating $L_{\nu}\propto\nu^{-\alpha_{ox}}$ 
from $\lambda=1216{\rm \AA}$ to 1 keV.
The slope of the EUV SED ($\alpha_{ox}$) is given by Eqn.\ 2 of \citet{vign03},
\begin{equation}\label{eqAlphaOx}
\alpha_{ox}=0.1\log\left[\frac{L_{\nu}(2500{\rm \AA})}{erg\ s^{-1}}\right]-1.32
\end{equation}
with $L_{\nu}(2500{\rm \AA})$ taken from the AGN template SED.
Finally, we eliminate reprocessed emission from dust by
assuming $F_{\nu}\propto\nu^{-2}$ for $1\mu m<\lambda<30\mu m$, following
M04.  

\begin{figure}
\epsscale{1.0}
\plotone{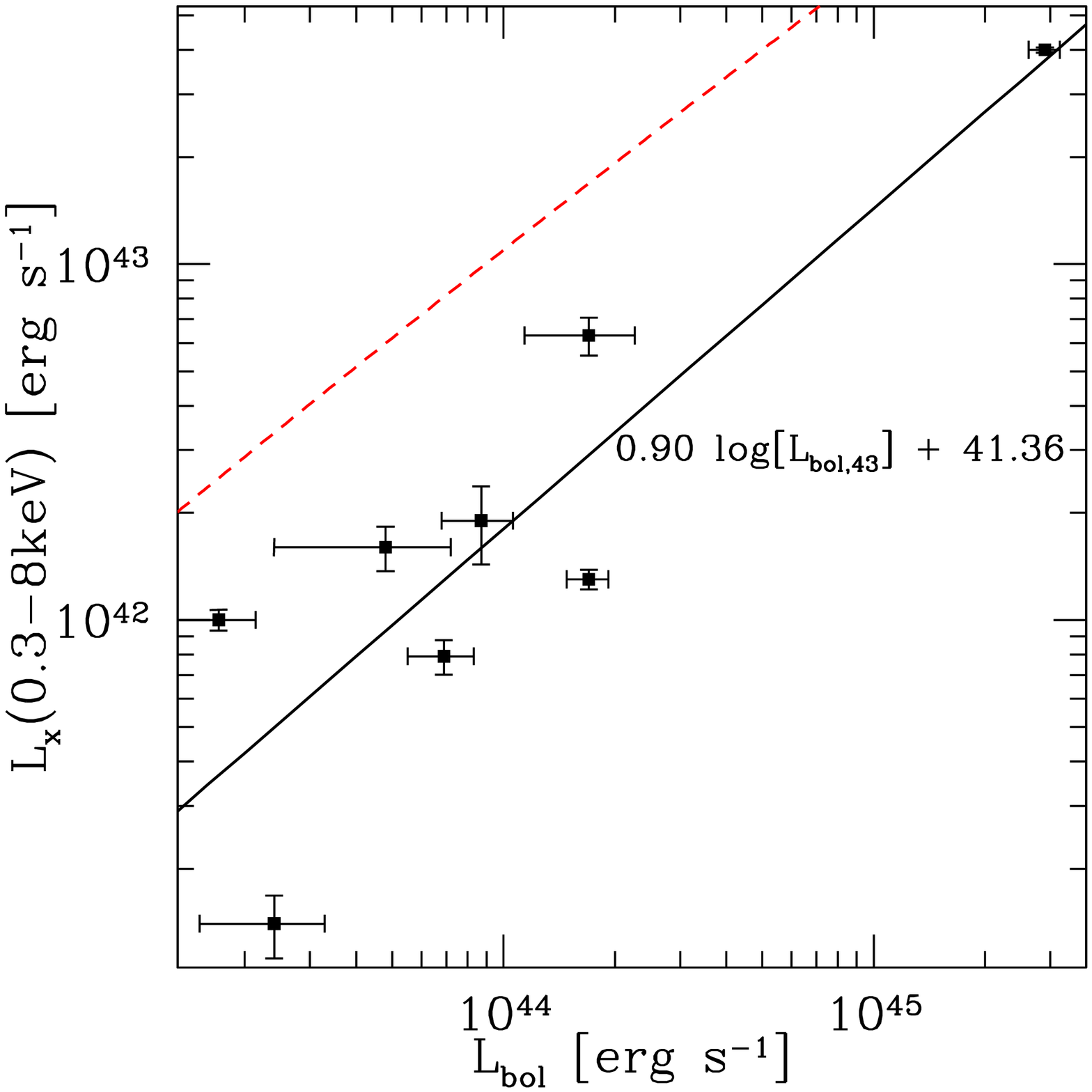}
\caption{Comparison of our estimated relation between $L_{bol}$ and
$L_{\rm X}$ ({\it black solid line}) with that found by M04
({\it red dashed line}).  The measured X-ray (0.3--8 keV) and
bolometric luminosities of AGNs identified by both the X-ray
and IR selection criteria are shown by {\it black points}.
$L_{bol}$ is derived by integrating
the A10 AGN model SED component from $1216{\rm \AA}$ to $1\mu m$
and assuming a declining continuum with $F_{\nu}\propto\nu^{-2}$ for
$\lambda>1\mu m$.  $L_{\rm X}$ is determined assuming
a $\Gamma=1.7$ power law.
See Section \ref{secXraySens} for further information.
\label{figBC}}
\end{figure}

To correct the X-ray luminosities of X-ray AGNs to bolometric luminosities,
we fit a power-law to the measured $L_{\rm X}(0.3-8\ {\rm keV})$
and $L_{bol}$ of the 8 IR AGNs identified separately in X-rays.
A least-squares fit to the total X-ray and AGN luminosities yields,
\begin{align}\label{eqBC}
&\log[L_{\rm X}(0.3-8\ {\rm keV})]=(0.9\pm0.2)\log\biggl[\frac{L_{bol}}{10^{43} {\rm erg\ s^{-1}}}\biggr] \nonumber \\
&+(41.4\pm0.2)
\end{align}
where $L_{bol}$ is the bolometric AGN luminosity integrated from
10 keV to $30\mu m$. 
The AGNs used to determine Eqn.\ \ref{eqBC}
show a scatter of 0.4 dex about the best-fit relation 
(Figure \ref{figBC}).  Figure \ref{figBC} suggests that the slope
returned by the fit may be strongly influenced by the highest-luminosity
AGN.  However, a fit to the other 7 AGNs returns an identical
slope ($0.9\pm0.5$), so Eqn.\ \ref{eqBC} is not significantly biased by 
the highest-luminosity
object.  The luminosity dependence of the bolometric corrections (BCs)
derived from
the fit is therefore robust.  The slope is also consistent,
within large statistical
uncertainties, with the luminosity-dependence derived by M04.

The BCs derived from Eqn.\ \ref{eqBC} are fairly crude.  For
example, the fit does
not account for uncertainties on $L_{\rm X}$ or $L_{bol}$.
It also ignores upper limits, which will lead it
to over-predict the true $L_{\rm X}$ at fixed $L_{bol}$.
M04, by contrast, provide luminosity-dependent
BCs in several energy ranges that account for X-ray non-detections
(their Eqn.\ 21).
We convert their BCs to 0.3--8 keV assuming $\Gamma=1.7$ and
estimate the expected X-ray flux from our IR AGNs.  The predicted
X-ray fluxes exceed those estimated using Eqn.\ \ref{eqBC}, which we
know over-estimates the intrinsic $L_{\rm X}$--$L_{bol}$ relation, by
0.7 dex or more.  This might result if the M04 SED is a poor
match to the A10 AGN template.  M04 determine their X-ray BCs
using the $\alpha_{ox}$ relation derived by \citet{vign03} for
a sample of SDSS quasars, including broad-absorption line quasars
(BALQSOs).  Given that our $L_{bol}$ calculation is insensitive to the
absorption in BALQSOs, it is possible that the M04 BCs
over-estimate $L_{\rm X}$ at fixed $L_{\nu}(2500{\rm \AA})$
when applied to our sample.
In order to produce consistent results for the X-ray and
IR AGNs, we therefore use the BCs implied by Eqn.\ \ref{eqBC} rather than the
M04 BCs, despite the large uncertainties associated with Eqn.\ \ref{eqBC}.

\subsection{X-ray Sensitivity}\label{secXraySens}
With the $L_{\rm X}$--$L_{bol}$ relation provided by Eqn.\ \ref{eqBC},
we can determine whether the X-ray non-detection of many IR AGNs results
from some intrinsic difference between the two classes of AGNs or
if it is merely a result of the sensitivity of the X-ray images used by M06.
Eqn.\ \ref{eqBC} predicts that 9 (5) IR AGNs with no X-ray detections
should be more than a factor of 3 (5) brighter
than the faintest point source in their parent clusters (M06).
The M04 BCs produce more X-ray flux at fixed bolometric luminosity
than Eqn.\ \ref{eqBC} and yield 13 (12) IR AGNs with significant X-ray
non-detections with the same flux limits.  The lack of detectable X-rays
from many IR AGNs is consequently easier to explain if we use Eqn.\
\ref{eqBC} rather than the M04 relations to predict their X-ray luminosities.

The minimum detected flux in a given cluster may not always be a fair
representation of the sensitivity for a given IR AGN due
to variations in the {\it Chandra} effective area with 
off-axis angle.  However,
the magnitudes by which many IR AGNs in AC 114 exceed the minimum detected
flux, sometimes more than a factor of 5, suggest that these AGNs
should have been detected if they obeyed the
$L_{\rm X}$--$L_{bol}$ relation of Eqn.\ \ref{eqBC}.  The non-detection
of many IR AGNs in X-rays
is qualitatively consistent with the results of \citet{hick09},
whose IR AGN selection relied upon the Stern wedge, and who found
many strong IR AGNs that could not be identified in X-rays.  At least some of
the ``missing'' IR AGNs could be highly obscured.  An intervening absorber
with $N_{H}=10^{22}\ {\rm cm}^{-2}$ would reduce the observed 0.5-2 keV
flux by a factor 3, which is sufficient to explain many of the
missing IR AGNs.  The missing AGNs could also result 
from the large scatter about the
mean $L_{\nu}(2500{\rm \AA})$--$\alpha_{ox}$ relation.  The AGN with the most
significant X-ray non-detection exceeds the minimum reported flux by a factor
of 7, which can be explained by $\Delta\alpha_{ox}\approx0.4$.
\citet{vign03} report a large intrinsic scatter about
their best-fit relation, and the combination of this scatter with
{\it in situ} absorption could mask moderately luminous
AGNs from detection in X-rays.

Finally, at least one IR AGN (A1689 \#109) appears to be
absent from the M06 sample due to X-ray variability
rather than as a result of absorption, intrinsic X-ray faintness, or
shallow Chandra imaging.  This object is moderately
luminous ($L_{bol}=2.1\times10^{10} L_{\odot}$), AGN-dominated 
($f_{AGN}=0.95$), falls firmly in the
middle of the Stern wedge, and is very robustly detected
by our likelihood ratio selection ($\rho=4\times10^{-77}$).
Nevertheless, there is no X-ray point source associated with this
object in the {\it Chandra} image employed by M06.  In a
more recent observation ({\it Chandra} Obs ID 6930, PI G. Garmire),
A1689 \#109 is associated with an X-ray point source
far brighter than the X-ray sources reported by M06.
It therefore seems likely that the IR AGNs that require
the most extreme values of $\alpha_{ox}$ could be accounted for by
variability rather than by systematically weak X-ray emission
compared to their visible-wavelength luminosities.

\subsection{Host Galaxies}\label{secHosts}
We determine stellar masses and SFRs for AGN host galaxies 
after subtracting the AGN component from the SED.  This introduces
some additional uncertainty in the resulting masses and SFRs beyond
the original photometric uncertainties, as discussed in Section
\ref{secMass}.  The uncertainty in the AGN
contribution to the measured MIR fluxes 
can prevent detection of low-level star
formation in IR AGNs.  The SFR distribution among IR AGNs
is therefore biased toward high SFR.

\begin{figure}
\epsscale{1.0}
\plotone{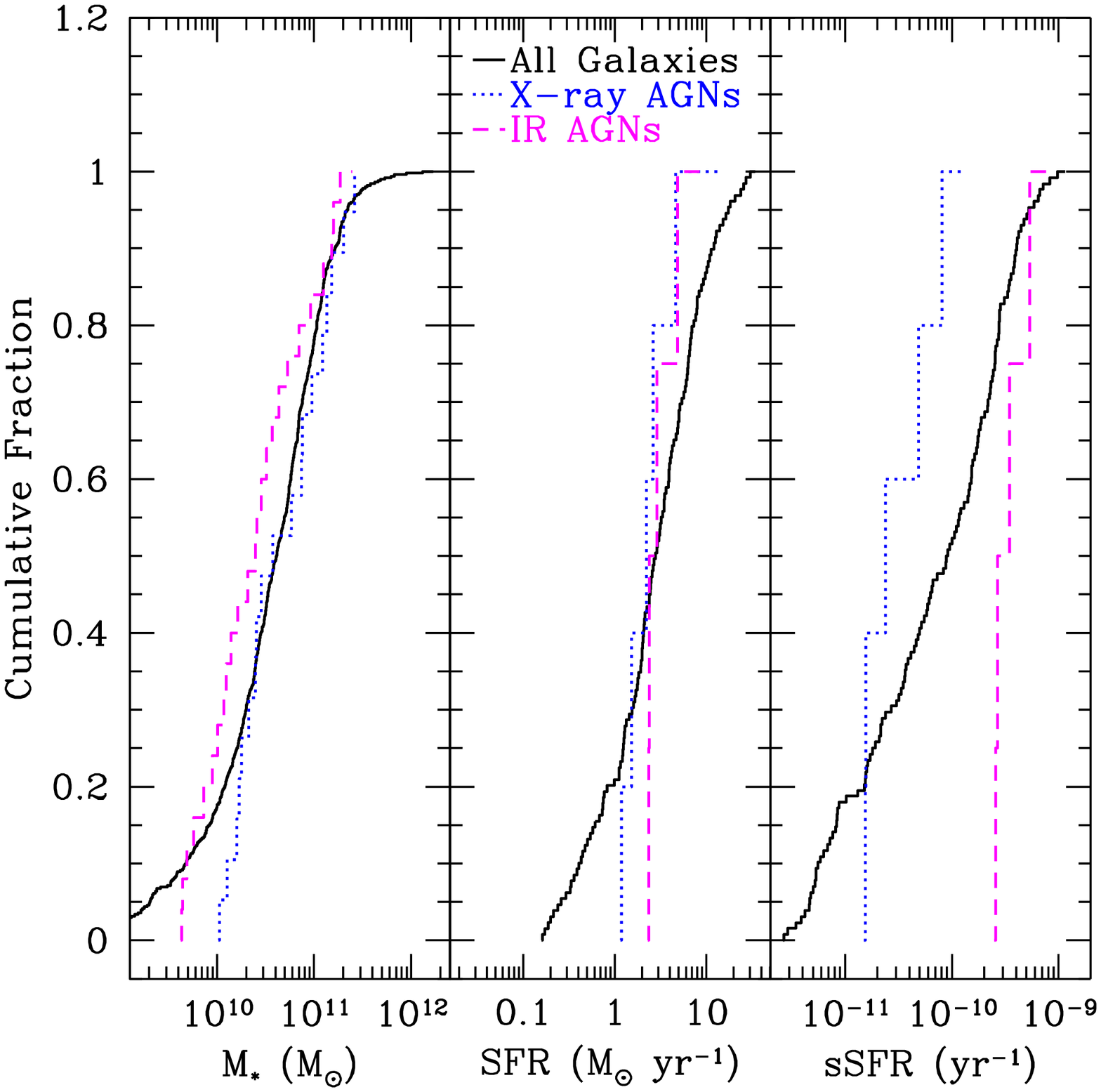}
\caption{Cumulative stellar mass, SFR and sSFR distributions of the 
X-ray and IR AGN samples compared to the distributions for all
cluster members.  Neither of the AGN samples show any significant
differences in either $M_{*}$ or SFR compared to the full sample
of cluster members, nor does the merged AGN sample.
However, IR AGN hosts have higher sSFR than both
X-ray AGN hosts and normal galaxies at 99\% confidence,
despite their similarities in $M_{*}$ and SFR.
\label{figHostCompare}}
\end{figure}

Figure \ref{figHostCompare} shows the results of comparing galaxies
hosting different types of AGNs to one another and also to
cluster galaxies as a whole.  The stellar mass and SFR
distributions of galaxies hosting X-ray and IR AGNs show
no measurable differences
with the distributions of all cluster members.  Merging the X-ray and IR
AGN samples likewise yields no measurable difference.
However, the hosts of IR AGNs have high specific SFRs 
(sSFR) compared to the hosts of
X-ray AGNs and to all cluster members
at 98\% and 97\% confidence, respectively.  The
difference between the sSFRs of X-ray AGN hosts and the full galaxy
sample is not significant.  However, X-ray AGN hosts appear to have
lower sSFRs than the average galaxy in Figure \ref{figHostCompare},
which is consistent
with previous results using field galaxies \citep{hick09}.
We must also consider the effect of non-detections on
the measured distributions.  Many of the IR AGN
hosts have upper limits on SFR that are smaller than the 
SFRs of the X-ray AGN host galaxies with the lowest
measurable SFRs.  Therefore, if the IR AGN hosts had
a distribution of SFRs similar
to the X-ray AGN hosts with measurable star-formation,
star-formation would have been detected in most IR AGN hosts.
This indicates that uncertainties in the AGN corrections alone
cannot account for the higher sSFRs among IR AGN hosts.

The IRAC color-color diagram (e.g.\ \citealt{ster05})
probes of the nature of AGN host galaxies independent of their
model SEDs by identifying
the dominant source of their MIR emission \citep{donl08}.
The MIR colors of X-ray and IR AGNs before their AGN
components are subtracted are compared to all cluster
members in Figure \ref{figStern}.  Galaxies hosting AGNs
have unremarkable $[5.8]-[8.0]$ colors but do not extend as far
to the red as normal galaxies, which indicates that AGNs are seldom
found in starbursts or luminous infrared galaxies
\citep{donl08}.  AGN hosts also show redder
$[3.6]-[4.5]$ colors than typical for a red sequence galaxy, which may
indicate a contribution of hot dust to the $4.5\mu m$ continuum.
The colors of AGN hosts, especially IR AGN hosts, are
influenced by the AGN continuum, but tests using the AGN and spiral galaxy
templates indicate that only galaxies in
the Stern wedge have more than 50\% of their IRAC fluxes contributed
by the AGN component.
A two-dimensional KS-test confirms that, after excluding objects
in the Stern wedge, the IRAC colors of
both X-ray and IR AGNs differ from normal galaxies
at $>99.9\%$\ confidence,
and the absence of X-ray AGNs among the most vigorously
star-forming galaxies (those with the reddest
$[5.8]-[8.0]$ colors) is consistent
with earlier indications that X-ray AGNs avoid the
blue cloud in visible color-magnitude diagrams (CMDs; 
\citealt{scha09,hick09}).  The distribution of X-ray
AGNs in Figure \ref{figStern} also appears to be consistent with the results
of \citet{gorj08}, who found that $16.8\pm0.3\%$
of X-ray--identified AGNs outside
the Stern wedge had very red $[5.8]-[8.0]$ colors consistent with
vigorous, on-going star-formation.  We found this population to be 
$20\pm6$\% among our X-ray AGNs. 

\begin{figure}
\plotone{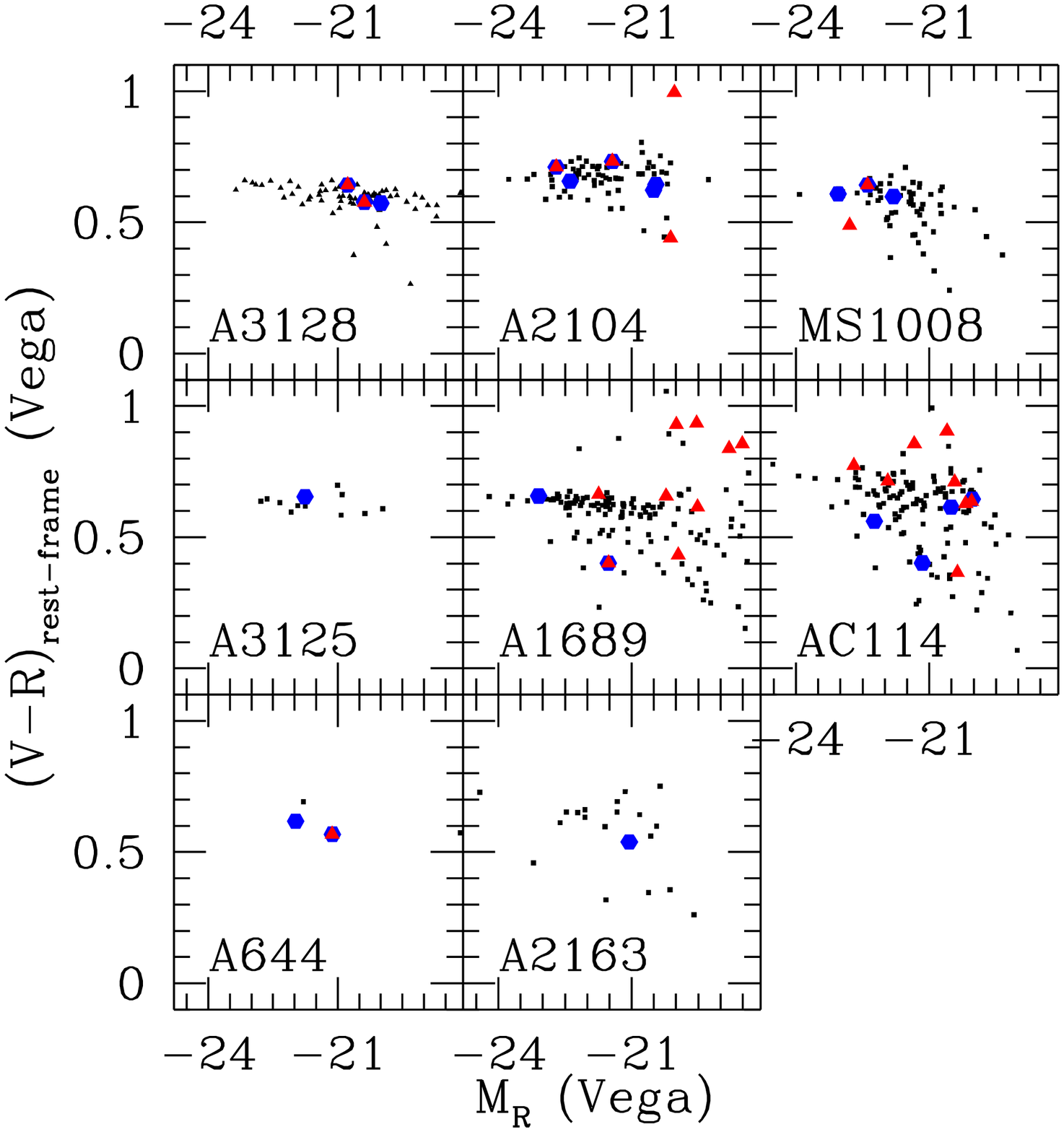}
\caption{Optical color-magnitude diagrams showing the
spectroscopically-confirmed member galaxies
({\it black squares}), X-ray AGNs ({\it blue hexagons}) and IR AGNs
({\it red triangles}) of each cluster.  The contribution of the AGN
component to the model SED has been subtracted from the IR AGNs,
leaving estimated host-galaxy colors and luminosities.  Typical
uncertainties on colors and absolute magnitudes are approximately
0.1 mag.  See Section
\ref{secHosts} for further discussion.
\label{figColorMag}}
\end{figure}

The visible CMD provides a means to estimate the nature of galaxies
in the absence of measurable star-formation at MIR wavelengths.  Figure
\ref{figColorMag} shows the CMD for each cluster after
the AGN component has been subtracted.
The fraction of cluster members hosting an X-ray AGN peaks
on the red sequence, and the probability that the X-ray AGN hosts
are drawn from the parent cluster population is less than $10^{-3}$.
This contrasts with AGN hosts in the field, where the X-ray AGN fraction
typically peaks in the green valley (\citealt{hick09};
\citealt{scha09}; \citealt{silv09}, henceforth S09) 
for AGNs identified using either
X-ray luminosity or emission-line diagnostics.
IR AGN hosts, both in our sample of cluster AGNs and in the field
sample of \citet{hick09}, conspicuously avoid the red sequence.
Like the difference between X-ray AGN hosts and the parent cluster
population, this result is significant at $>99.9\%$ confidence.
This indicates that the IR AGN sample has at most limited contamination by
MIR-excess early-type galaxies of the sort studied by, e.g.\ \citet{bran09}.
Galaxies hosting IR AGNs in clusters also show an important difference
compared to their counterparts in the field.  While only 1.5\% of field galaxies
hosting the IR AGNs studied by \citet{hick09} had $^{0.1}(u-g)$ colors
redder than the median of the red sequence,
more than 20\% of IR AGNs in clusters have visible
colors redder than the red sequence in their parent clusters.

We examined the SDSS $g-r$ colors of very red galaxies 
$\bigl((V-R)_{rest-frame}>0.8\bigr)$
in Abell 1689, which has the largest number of such objects, and found
that most also appear red in SDSS colors.
The most notable exception is Abell 1689 \#192, which we have identified
as an IR AGN, which suggests that its colors may change due to 
AGN variability.
The qualitative agreement between the colors of very red galaxies in
Figure \ref{figColorMag} and their $g-r$ colors from SDSS
suggests that these objects
are genuinely unusual and not the result of photometric errors.
These galaxies also show substantial reddening of the AGN template in their
A10 fit results, with $\langle E(B-V)\rangle =0.4$ and a 
trend for higher $E(B-V)$ in galaxies with redder colors
at 97\% confidence.  These results suggest that 
the unusually red galaxies in Figure
\ref{figColorMag} experience significant internal extinction that
is not present in most galaxies.

Since the AGN
component of the SED fit may account not only for a true AGN contribution
but also for intrinsic variations about the normal
galaxy templates, some or all of these very red AGNs, which represent
approximately 1/3 of our IR AGN sample, may not be true AGNs.
However, fewer than
half (7/17) of objects with $(V-R)_{rest-frame}>0.8$ are identified 
as IR AGNs; this implies that IR AGNs must
differ from normal galaxies not only in the visible but also in the MIR,
and MIR fluxes are practically immune to extinction.
Therefore, most of the IR AGNs identified in this region of 
color-magnitude space cannot be false-positives selected
due to their unusual visible
colors but must have genuine nuclear activity contributing to their SEDs.

\subsection{Accretion Rates}\label{secAccretion}
We use the bolometric luminosities of both X-ray and IR AGNs to
measure the growth of their black holes and compare the black hole
growth to the assembly of stellar mass in their host galaxies.
The accretion rate of a black hole can be generically written as,
\begin{equation}\label{eqAccretion}
\dot{M}_{BH} = \frac{L_{bol}}{\epsilon c^{2}}
\end{equation}
where $L_{bol}$ is the bolometric luminosity and
$\epsilon$ is the efficiency of conversion between the rest mass energy
($\dot{M}c^{2}$) of the accreted material and the energy radiated by the black
hole.  We assume $\epsilon=0.1$, appropriate for a thin accretion disk around
an SMBH with moderate spin \citep{thor74} and determine $L_{bol}$
as described in Section \ref{secBol}.

\begin{figure}
\epsscale{1.0}
\plotone{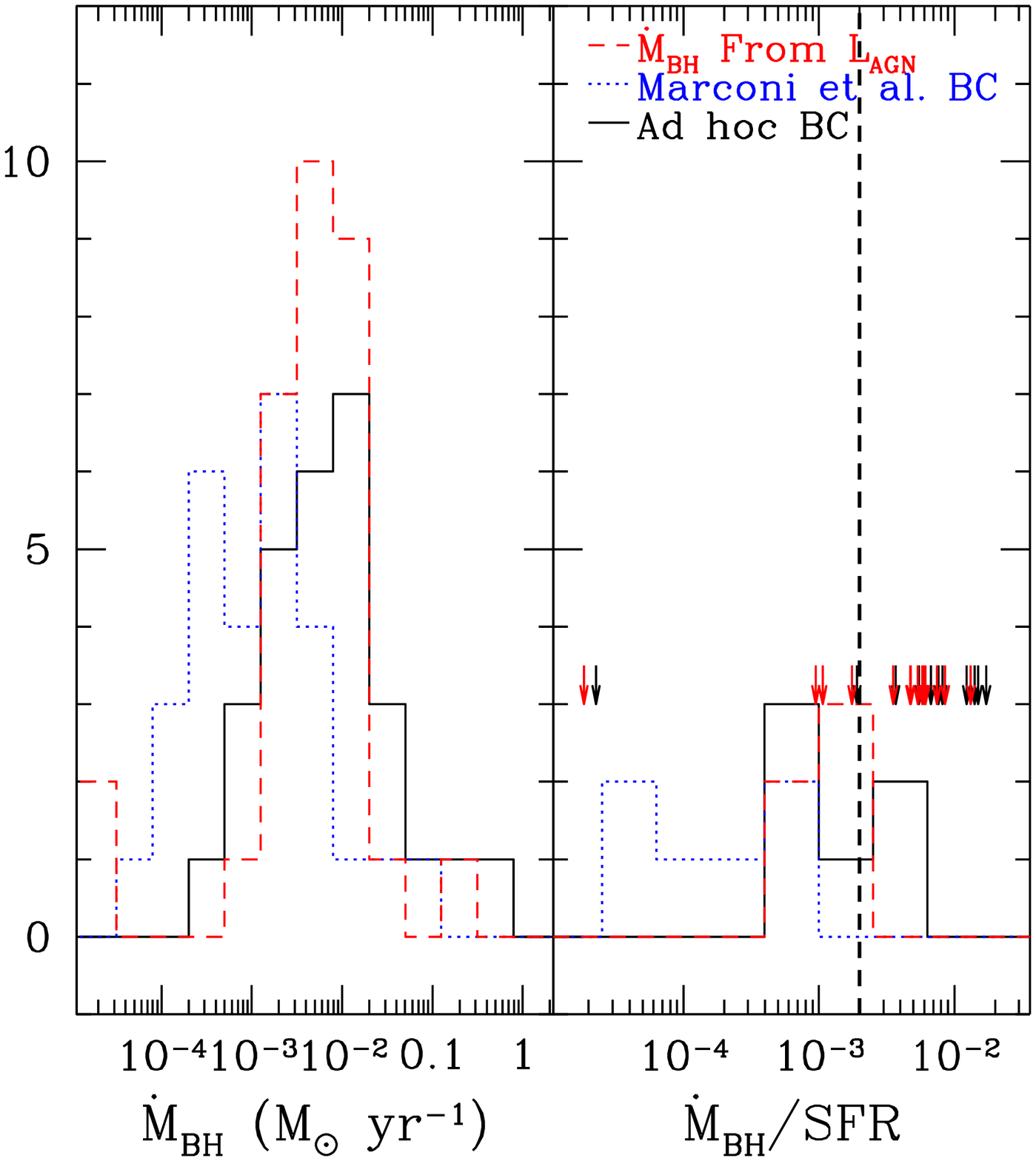}
\caption{Comparison of accretion rates ($\dot{M}_{BH}$; left panel) and
SMBH growth rates relative to their host galaxies 
($\dot{M}_{BH}/SFR$; right panel).  The various distributions
compare different methods of estimating $\dot{M}_{BH}$:
directly from the SEDs of IR AGNs ({\it red dashed}), applying
{\it ad hoc} BCs derived from AGNs with both X-ray and IR identifications
({\it solid black}), and applying the M04 BCs
({\it blue dotted}).  M04 BCs return significantly
lower $\dot{M}_{BH}$ than the other two methods, which
are consistent with one another.  Short {\it black}
and {\it red} arrows mark upper limits for X-ray and IR AGNs,
respectively.  The {\it dashed vertical line} marks the ratio required to
maintain the $z=0$ $M_{BH}$--$M_{bulge}$ relation \protect\citep{marc03}.
See Section \ref{secAccretion}
for more on the various BCs.
\label{figGrowth}}
\end{figure}

The accretion rates derived from Eqn.\ \ref{eqAccretion} for the
X-ray and IR AGN samples are shown in Figure \ref{figGrowth}.
The left panel suggests that X-ray and IR AGNs have similar
accretion rates, and a KS test reveals that there is no
significant difference between the two samples.
This is surprising, since we would
na\"ively expect that the difference between X-ray and IR AGNs
might be due to different dependence of X-ray and IR selection
techniques on luminosity.
Instead, the right panel of Figure \ref{figGrowth}
shows that the X-ray and IR AGN samples have 
$\langle \dot{M}_{BH}/SFR\rangle =3\times10^{-3}$ and 
$\langle \dot{M}_{BH}/SFR\rangle =2\times10^{-3}$,
respectively.  These ratios are comparable to the mean 
$M_{BH}/M_{bulge}$ in the local universe
($2\times10^{-3}$, \citealt{marc03}), which indicates that
the SMBHs in cluster AGNs are accreting at approximately the
rate required to maintain the $z=0$ $M_{BH}$--$M_{bulge}$ relation.
However, this is likely an artifact of our SFR detection thresholds,
as the accretion rates of these objects are not large enough to
produce outliers on the $M_{BH}$--$M_{bulge}$ relation in a Hubble time.

\begin{figure}
\epsscale{1.0}
\plotone{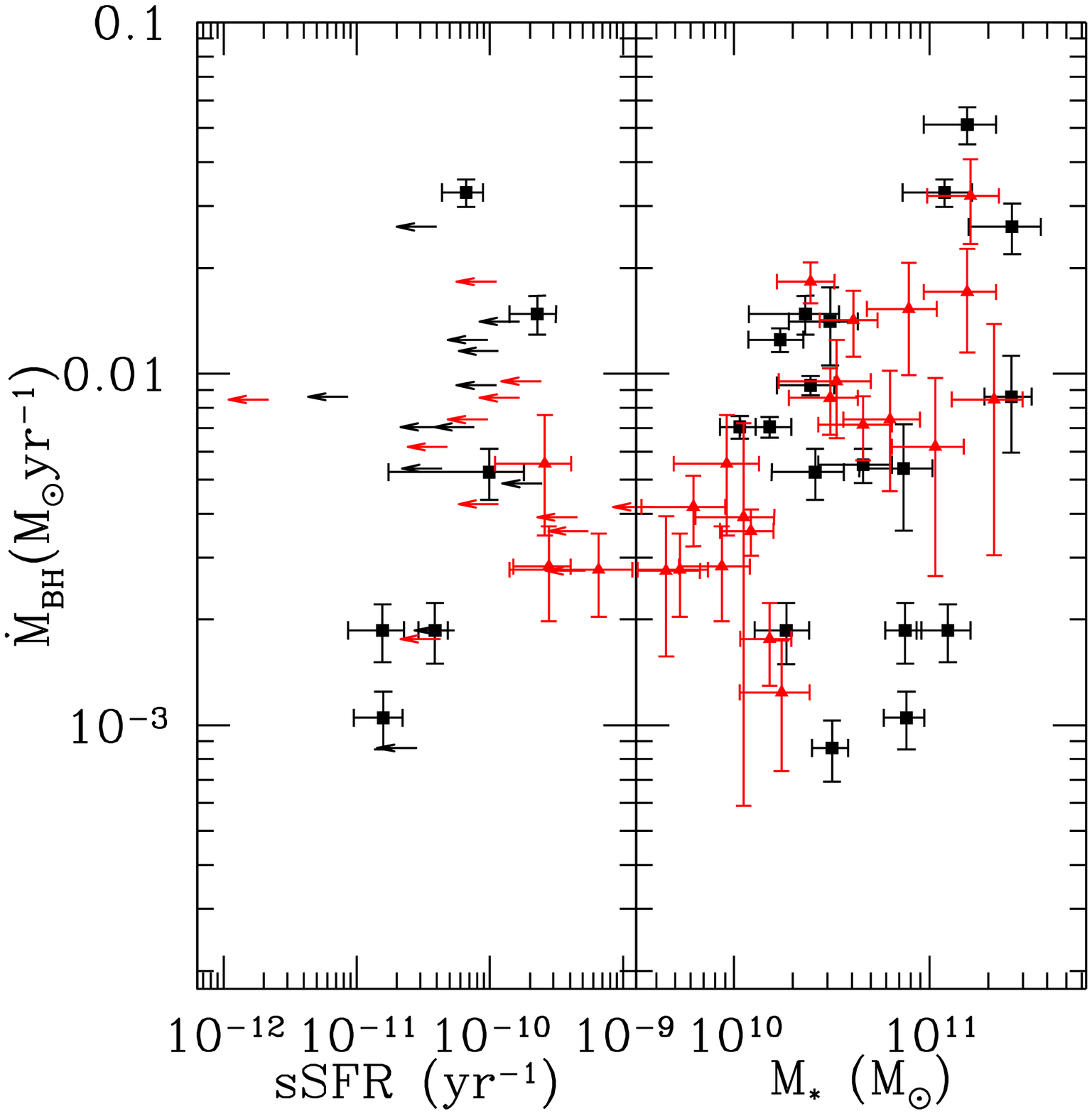}
\caption{Relationships of black hole accretion rates ($\dot{M}_{BH}$)
to stellar masses and sSFRs.  Black points and arrows show
$\dot{M}_{BH}$ inferred from X-ray luminosities
using BCs from Eqn.\ \ref{eqBC}, while red points and arrows
mark $\dot{M}_{BH}$ inferred from integrating model SEDs.  
Stellar masses and SFRs are measured from SEDs
and include the entire galaxy, not just the spheroidal
component.
\label{figCorrelations}}
\end{figure}

Figure \ref{figCorrelations} compares black hole accretion rates
with host mass and sSFR.  We find no significant correlation between
$\dot{M}_{BH}$ and sSFR, nor do we find a
correlation of $\dot{M}_{BH}$ with stellar mass
among the X-ray AGN sample.  However, $\dot{M}_{BH}$ correlates with
stellar mass among IR AGNs at 99.5\% confidence, weaking to 98\%
confidence among the merged AGN sample.  This correlation may be
related to the ability of more massive cluster members to retain
more cold gas.

\begin{figure*}
\epsscale{1.0}
\plotone{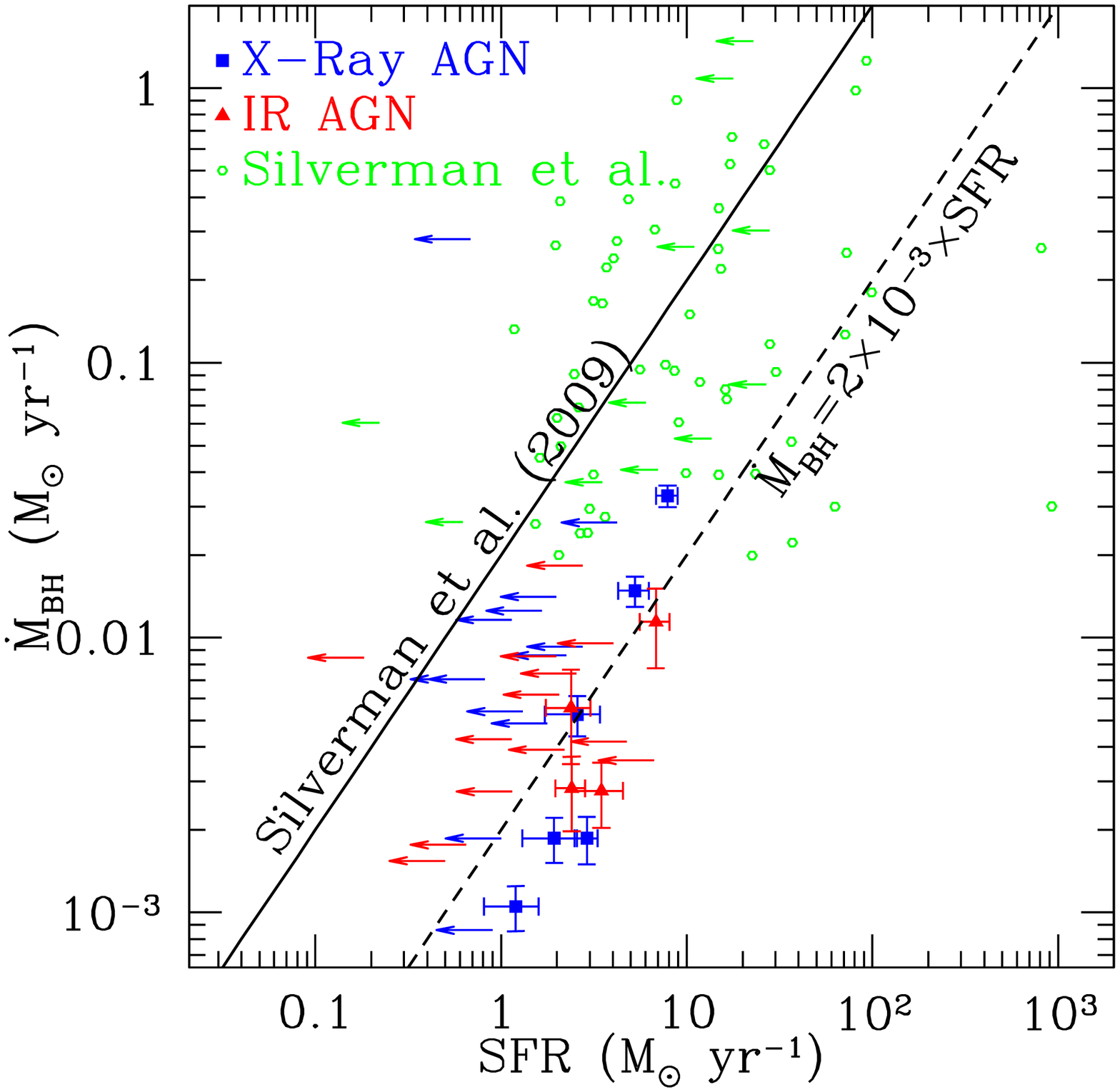}
\caption{Relationship between black hole growth and star-formation
in our AGN sample compared to the AGN sample from COSMOS
examined by S09.  Green circles and arrows mark
the S09 AGNs.  All other symbols are the same as in Figure
\ref{figCorrelations}.
Lines mark the $\dot{M}_{BH}/SFR$ relation
measured by S09 ({\it solid}) and the ratio required to
produce the $z=0$ $M_{BH}$--$M_{bulge}$
relation (\protect\citealt{marc03}; {\it dashed}).  Our SFRs and those
reported by \protect\citet{silv09} are galaxy-wide SFRs rather than bulge
SFRs.
\label{figCorrSfr}}
\end{figure*}

Figure \ref{figCorrSfr} shows the relationship between black hole
growth and stellar mass assembly in AGN host galaxies.
The correlation
of $\dot{M}_{BH}$ with SFR
is extraordinarily strong ($>99.9\%$\ confidence), and both X-ray and
IR AGNs appear to follow the same relation, with
${\rm SFR}\propto\dot{M}_{BH}^{0.46\pm0.06}$.
\citet{netz09} studied emission line selected AGNs from SDSS and also
found a tight correlation between SFR and AGN luminosity across nearly
5 dex in $L_{bol}$.  However, their SFR--$\dot{M}_{BH}$ relation
(${\rm SFR}\propto\dot{M}_{BH}^{0.8}$)
is steeper than ours at $5.7\sigma$.  Furthermore, \citet{lutz10}
performed a stacking analysis of X-ray identified AGNs at $z\sim1$
and found no measurable correlation of SFR with $L_{bol}$ for AGNs with
$L_{\rm 2-10 keV}<10^{44}\ {\rm erg\ s^{-1}}$.  However, the
millimeter-bright, optically luminous QSOs studied by \citet{lutz08}
appear to be consistent with both \citet{netz09} and \citet{lutz10}.
The qualitative similarity of our results to those of
\citet{netz09} and \citet{lutz10} suggests that we are seeing
the same underlying relationship.
That both X-ray, IR and emission-line selected AGNs appear to show
the same general trend toward higher $\dot{M}_{BH}$ in
hosts with higher SFR suggests that accretion rates in all
of these objects are set by the size
of the global cold gas reservoir.  Such a relationship is also
predicted theoretically as a result of large-scale dynamical
instabilities, which drive cold gas to the centers of galaxies where
it can be accreted \citep{kawak08,hopk10}.  However, the quantitative 
discrepancies between
the various observational signatures of star-formation and gas accretion
indicate that further work on the
relationship between these phenomena is needed.

Figure \ref{figCorrSfr} also compares star-formation
and black hole growth among our AGN sample 
with the median ratio found by S09 and
the ratio needed to maintain the $z=0$ $M_{BH}$--$M_{bulge}$ relation.
In some cases
$\dot{M}_{BH}/SFR$ falls more than a dex below the ratio reported by
S09 for field galaxies at $z\approx0.8$ and more than
0.3 dex below the rate needed to maintain the local 
$M_{BH}$--$M_{bulge}$ relation.  However,
if we consider AGN hosts with no measurable star-formation, the disagreement
in $\dot{M}_{BH}/\dot{M}_{*}$
between the cluster AGNs we measure and the field AGNs of S09
becomes far less pronounced.  The upper limits in Figure \ref{figCorrSfr}
fill in much of the empty space between the S09
median relation and the cluster AGNs with measurable star-formation,
but the fraction of
galaxies with $\dot{M}_{BH}/SFR<2\times10^{-3}$ is larger
in Figure \ref{figCorrSfr} than in Figure 13 of S09
(7/39 versus 9/67).  This difference grows (7/27) if we consider
only AGNs with $\dot{M}_{BH}<10^{-2}\ M_{\odot}\ {\rm yr^{-1}}$,
which is below the luminosity
limit of the S09 sample.  However, even the difference between the
low-luminosity subsample and the S09 result is not statistically
significant (90\% confidence).  \citet{silv09}
project the evolution in the median SFR of their AGN sample to
$z=0$ and find that it agrees with the SFRs measured
in Type 2 AGNs with $\log(L_{[OIII]})>40.5$ in SDSS.
The median $z=0.2$ SFR for the S09 AGN hosts is 
$SFR\approx0.5\ M_{\odot}\ yr^{-1}$, which is comparable to our
detection threshold.  As a result, the AGNs measured
in Figure \ref{figCorrSfr} are more comparable to a high-SFR
subsample of the S09 AGNs.  However, there is no significant
difference in the $\dot{M}_{BH}/SFR$ of high-SFR versus low-SFR
AGNs in S09.  We therefore concluded that the ratio of
$\dot{M}_{BH}$ to SFR our sample of low-z cluster AGNs
is consistent with the ratios observed in high-z AGNs in the field.

\subsection{Radial Distributions}\label{secDistr}
\begin{figure}
\epsscale{1.0}
\plotone{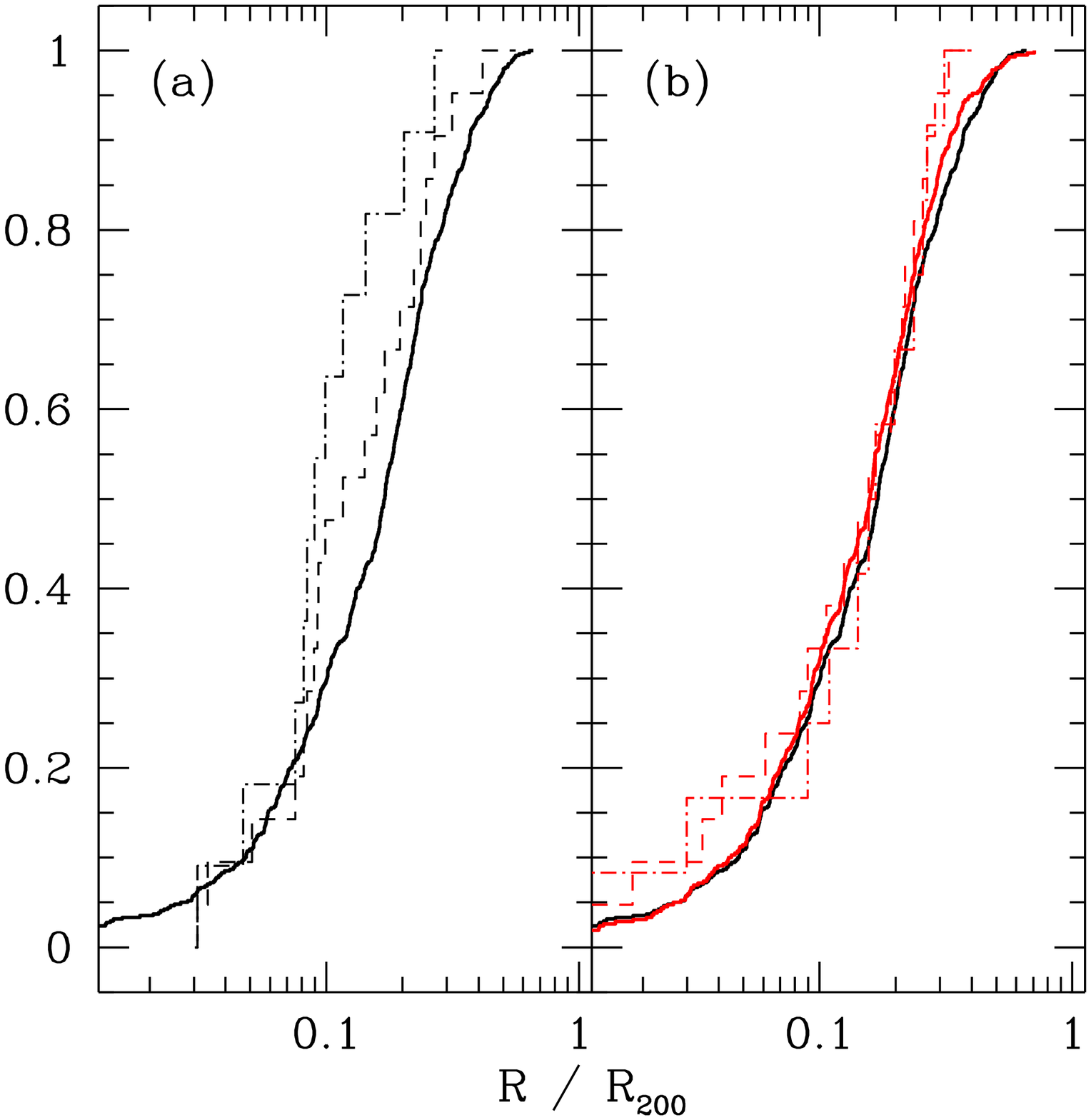}
\caption{Distributions of X-ray AGNs, IR AGNs and 
all cluster galaxies within their
parent clusters.  {\it Panel a}: luminous 
($L_{\rm X}>10^{42}\ {\rm erg\ s^{-1}}$; {\it dot-dashed})
and all X-ray AGNs ({\it dashed})
compared to all cluster members with $M_{*}>M_{cut}=3\times10^{10}\ M_{\odot}$
({\it solid}).
{\it Panel b}: luminous ($L_{bol}>10^{10}\ L_{\odot}$;
{\it dot-dashed red}) and all IR AGNs ({\it dashed red})
compared to cluster members
with {\it Spitzer} coverage and $M_{*}>M_{cut}$ ({\it solid red})
and the unbiased sample of galaxies inside the Chandra footprint
({\it solid black}).
\label{figRadial}}
\end{figure}

\citet{mart07} found that luminous ($L_{X}>10^{42}\ {\rm erg\ s^{-1}}$)
X-ray AGNs were more centrally concentrated in $R/R_{200}$ than
normal cluster members at 97\% confidence.
After pruning the AGN sample of suspect redshifts and applying
improved K-corrections,
we assemble the radial distributions of our AGN samples
in Figure \ref{figRadial}.
Figures \ref{figRadial}a and \ref{figRadial}b, which consider the
X-ray and IR AGN samples, respectively, have slightly different distributions
of parent galaxies.  This is because
{\it Spitzer} pointings cover only the fields around X-ray sources
identified by M06 and not the full {\it Chandra} field of view.
The IR AGNs are selected from the cluster member catalog after
SED fitting has been performed, so the radial distribution of IR AGNs
is guaranteed to be unbiased with respect to the cluster galaxy
sample we used above, while X-ray AGNs must be compared to the distribution
of all galaxies within the {\it Chandra} footprint.
These different selection footprints lead to the different
radial distributions shown in the solid red and black lines in
Figure \ref{figRadial}b.  The difference is not significant, however,
and has no impact on our conclusions.

We have determined that the host galaxy of the X-ray point source
identified as the cluster AGN AC114-5 by M06 
had an erroneous spectroscopic redshift reported
in the literature (see Section \ref{secModelSeds} and Figure 
\ref{figXrayFits}).  Our SED fitting indicates that this source is
a background QSO at $z_{phot}\approx0.99$.
Without this object, which is located at a projected distance
$R/R_{200}\approx0.2$ from the center of AC 114,
the significance of the difference between the luminous
X-ray AGN and control samples
drops to 89\% confidence with a luminosity-selected control sample
and 92\% confidence with a mass-selected control sample.
Consistent with the results of \citet{mart07}, we also find no
significant difference between the radial distribution of the
full X-ray AGN sample compared to the distribution of
cluster members as a whole.

Following \citet{mart09}, we also try a redshift-dependent 
luminosity threshold $\biggl(M_{R,cut}=M_{R}^{*}(0)+1-z\biggr)$
in place of a fixed value.  The galaxy and AGN samples selected using
this criterion show no significant differences in their $R/R_{200}$
distributions.
\citet{mart09} chose this evolving threshold to select a sample of
passively-evolving
galaxies at fixed stellar mass.  A mass threshold
($M_{*}>3\times10^{10}M_{\odot}$)
appropriate for an elliptical galaxy at $z=0$ with $M_{R}=M_{R,cut}$
again yields no measurable difference between the
radial distributions of X-ray AGNs and all cluster members.
We conclude that the radial distributions of both X-ray and IR
AGNs in galaxy clusters are consistent with the distribution
of cluster members, although the agreement between cluster
members and IR AGNs is much better than between cluster members
and X-ray AGNs.

\section{Discussion}\label{secDiscuss}
Identifying AGN from their X-ray emission is widely considered to be 
among the most robust means of selecting AGNs (e.g.\ \citealt{ueda03},
S09, A10), because the measured hard X-ray luminosity
of a given AGN is largely insensitive
to absorption if $N_{H}<10^{24}\ {\rm cm^{-2}}$.  Furthermore, the fraction
of Compton-thick AGNs ($N_{H}>10^{24}\ {\rm cm^{-2}}$) is
small, with 10\% or less of all cosmic black hole
growth taking place in Compton-thick systems \citep{trei09}.
Alternatively, AGNs can also be robustly identified from their
UV continuum emission after it has been absorbed by dust and re-emitted
in the MIR.  If these techniques are similarly immune to the effects
of absorption they should yield very similar AGN samples.
Instead, we find that at most 15\% of 
AGNs in galaxy clusters are identified by both X-ray and MIR techniques.

Furthermore, it is clear that this dichotomy does not result solely
from the relative luminosities of X-ray and IR AGNs.  The IR AGN
sample contains 5-9 objects that should have been detected in X-rays
if their SEDs were similar to those AGNs identified using both
selection methods.  The most prominent of these is Abell 1689 \#109,
which has $L_{bol}\approx8\times10^{45}\ {\rm erg\ s^{-1}}$ but was
not detected in the {\it Chandra} image used by M06
to identify X-ray AGNs in Abell 1689.  This AGN appears quite
prominently in a subsequent {\it Chandra} image, indicating that its initial
non-detection was most likely the result of X-ray variability.  This
example demonstrates that the absence of detectable
X-ray emission from an AGN candidate, even a fairly luminous one,
does not necessarily preclude the presence of an AGN.  However, Abell 1689
\#109 is not typical.  The IR AGNs with significant X-ray non-detections
are not necessarily the most luminous.  Instead, they reside
in the clusters with the
deepest X-ray images.  Indeed, all of the X-ray non-detections in AC 114
that fall within the {\it Chandra} image footprint are predicted to be
at least 3 times brighter than the faintest reported X-ray point source.
As a result, at least some of these non-detections could indicate
contamination of the IR AGN sample by one or more of the effects
discussed in Section \ref{secAgn}, e.g.\ intrinsic variation in
the AGN SED or dust heating by AGB carbon-stars.
More observational and theoretical work on the dust emission in old
stellar populations are required before the potential of these sources
of MIR emission to mimic an AGN-like SED can be quantified.

In the absence of detailed, calibrated models for ``contamination'' of
MIR emission by old stars, we assume that this component is
negligible.  This implies that X-ray selection alone can miss
a large fraction of moderate-to-low luminosity AGNs.  This could have
important implications for studies of star-formation in clusters
using MIR luminosities (e.g.\ \citealt{sain08,bai09,geac09}).
This is especially important if authors
assume that AGNs can always be identified with X-rays alone or
that the MIR emission from galaxies with X-ray excesses is always
dominated by AGN emission.  These assumptions
imply that any MIR emission not associated with an X-ray AGN must
be powered by star-formation and that no MIR emission from a galaxy
hosting an X-ray AGN can be powered by star-formation.
Our results indicate that these assumptions
may lead authors to overestimate the number of cluster galaxies
with vigorous star-formation and to underestimate the number with
moderate star-formation.
Therefore, additional tests for AGN are needed to correctly
interpret the MIR luminosities of cluster galaxies.

A difference between X-ray-- and MIR--selected AGN samples also
appears among field samples, which consist of more luminous AGNs than the
ones we study and use a different MIR selection method \citep{hick09}.
The color distributions of IR AGNs selected using different techniques
also differ from one another, but it is clear that galaxies hosting
AGNs identified from their X-ray emission are dissimilar from
galaxies hosting AGNs identified in the MIR.  Most notably, IR AGN
hosts have significantly higher sSFRs than the average cluster
galaxy, while there is no significant difference between
the sSFRs of X-ray AGNs and the cluster population as a whole.
Since SFR correlates well with cold gas mass, higher sSFRs
among IR AGN host galaxies suggests these galaxies have a larger
fraction of their baryons in cold gas than X-ray AGN hosts.  However,
the differences discussed
in Section \ref{secHosts} are determined only for galaxies with
measurable star-formation.  Several IR AGNs are found in host galaxies
that have both visible and IRAC colors consistent with passively-evolving
stellar systems.

The tight correlations between accretion rates of both X-ray and IR
AGNs with SFR in their host galaxies suggests that the two classes
are fueled by the same mechanism and are therefore fundamentally
similar.
Subject to the caveat described above, the larger sSFRs found in
IR AGN hosts might explain the apparent dichotomy of the two AGN classes
despite their physical similarity.
Larger gas fractions in IR AGN hosts
could lead to larger average column densities in IR AGNs,
depressing $L_{\rm X}/L_{bol}$ in these systems.
The presence of at least 5 of the 8 IR AGNs with X-ray counterparts
on the red sequence, where there is little cold gas to
participate in X-ray absorption, tends to support 
this scenario (Figure \ref{figColorMag}).  If cold gas fractions of AGN
host galaxies influence the detectability of X-ray AGNs, this
might also explain the dearth of X-ray AGNs
in the green valley in clusters compared to the field.  The X-ray AGNs in
our sample are weaker than the AGNs usually studied in field galaxy
samples, and a modest cold gas reservoir in green valley galaxies
could more easily absorb enough X-rays from an AGN with 
$L_{\rm X}=10^{41}\ {\rm erg\ s^{-1}}$ to make it undetectable.  Doing
the same for an AGN with $L_{\rm X}=10^{43}\ {\rm erg\ s^{-1}}$,
which is more typical for the field samples studied by, e.g.\ \citet{hick09}
and S09, would require a larger gas column.

Just over half (58\%) of the M06 X-ray point sources have
detectable hard X-ray emission, and therefore many AGNs near the
{\it Chandra} detection limits could be hidden by a sufficiently
large absorbing column.  Only 3 of the 9 IR AGNs in AC114 whose bolometric
luminosities imply that they should have been detected in X-rays,
but were not, would remain detectable in the soft X-ray band behind
a gas column with $N_{H}=10^{22}\ {\rm cm^{-2}}$.
This column density is large for Type I AGNs, but it is
not unusual for Type II AGNs observed in X-rays \citep{ueda03}.
Furthermore, X-ray and IR AGNs seem to obey the same relationship
between SFR and accretion rate in AGN hosts whose SFRs are
measurable.  This is consistent with the hypothesis that 
the apparent dichotomy between X-ray and IR AGNs is false,
and the shape of an AGN's SED depends strongly on the amount of
absorbing material between us and the central black hole.

The scenario we propose, in which absorption by cold gas in the host 
galaxy is responsible for the absence of detectable X-ray emission from 
IR AGNs, is consistent with the differences we find between the two
samples.  However, verifying that absorption by the host ISM
is indeed the cause of this observed difference will 
require deeper X-ray observations to detect X-ray counterparts
and estimate absorption columns.
If this can be accomplished, the presence
of spectral signatures of X-ray absorption would confirm that the
host galaxy is responsible for hiding some IR AGNs from X-ray detection.

\section{Conclusions}\label{secConclusion}
We have used {\it Spitzer} imaging of galaxy clusters to identify AGNs
and to measure the masses and star-formation rates of their host
galaxies.  We find that AGNs identified by this technique have very little
overlap with AGNs identified in X-rays.
We compared the host galaxies of AGNs identified using the two
methods and determined that, while their masses and SFRs are 
indistinguishable, IR AGNs reside in galaxies with higher sSFRs
than both X-ray AGN hosts and the parent sample of cluster galaxies.
The hosts of X-ray AGNs have sSFRs that are somewhat
lower than but consistent with the sSFRs
seen in cluster galaxies as a whole.
The difference between X-ray AGN hosts
and normal cluster galaxies is significant only when comparing their
positions in visible color-magnitude and MIR color-color diagrams.
X-ray AGN hosts are rarely found in the regions of both
diagrams associated with vigorous star-formation.

We also find that accretion rates of both X-ray and IR AGNs correlate
strongly with SFR in their host galaxies.  This suggests that X-ray
and IR AGNs are physically similar and are fueled by the same
mechanism.
We hypothesize that the larger sSFRs seen in IR AGN hosts indicate
larger cold gas fractions in these galaxies, and suggest that this
could account for the apparent dichotomy between X-ray and IR AGNs.
A moderately large cold gas column density of $10^{23}\ {\rm cm^{-2}}$
could suppress the X-ray emission from the IR AGNs enough that we would
be unable to detect them.  The presence of IR AGNs but not X-ray AGNs
in galaxies with very red optical colors, indicative of strong
absorption, lends credence to this hypothesis.
It might also be verifiable directly by deep X-ray observations
of either AC 114 or Abell 1689 to search for X-ray emission from IR
AGNs and to determine if such X-ray emission shows evidence
for absorption intrinsic to the host galaxy.  For example, the most luminous
IR AGN with no X-ray counterpart in Abell 1689 could be detected by Chandra
with ${\rm S/N}=3$ per resolution element
at $4\ keV$---the energy cutoff for
objects with $N_{\rm H}=10^{23}\ {\rm cm^{-2}}$---in $160\ ks$.  This would
allow a crude model spectrum to be constructed and the intrinsic 
absorption column to be measured.  Finally, we have obtained NIR spectra
of several IR AGN in Abell 1689, which we will examine for high-ionization
emission lines that would unambiguously indicate the presence of an AGN.

Following \citet{mart07}, we compared the radial distributions of
AGNs and all cluster members.  We eliminated one
AGN with a spectroscopic redshift from the literature that incorrectly 
identified a background quasar as
a cluster member.  Without this object, the significance
of their result that
luminous X-ray AGNs ($L_{\rm X}>10^{42}\ {\rm erg\ s^{-1}}$) are
more concentrated than cluster members as a whole is reduced to
$\sim90\%$\ confidence.  While this result is no longer significant,
it would be worthwhile to extend the present sample
using archival {\it Chandra} imaging of additional clusters to
either confirm or refute that X-ray luminous AGNs are more concentrated
than the galaxy populations of their parent clusters.
It is unlikely, however, that a similar exercise using IR AGNs
would yield a positive result, as the radial distribution of IR AGNs
agrees very closely with the distribution of cluster galaxies.

\acknowledgements
We are grateful to John Silverman for providing his data.
We also thank Chris Kochanek for insightful comments on an earlier
draft.  PM is grateful for support from the NSF via award AST-0705170.
DWA thanks The Ohio State University for support from the Dean's Distinguished
University Fellowship.
This work is based in part on observations made with the Spitzer Space
Telescope, which is operated by the Jet Propulsion Laboratory, California
Institute of Technology under a contract with NASA. Support for this work
was provided by NASA through an award issued by JPL/Caltech.  
This research has made use of the NASA/IPAC Extragalactic Database (NED)
which is operated by the Jet Propulsion Laboratory, California Institute of
Technology, under contract with the National Aeronautics and Space 
Administration.

\clearpage
\begin{figure*}
\includegraphics[bb= 0 0 612 792,width=0.99\textwidth]{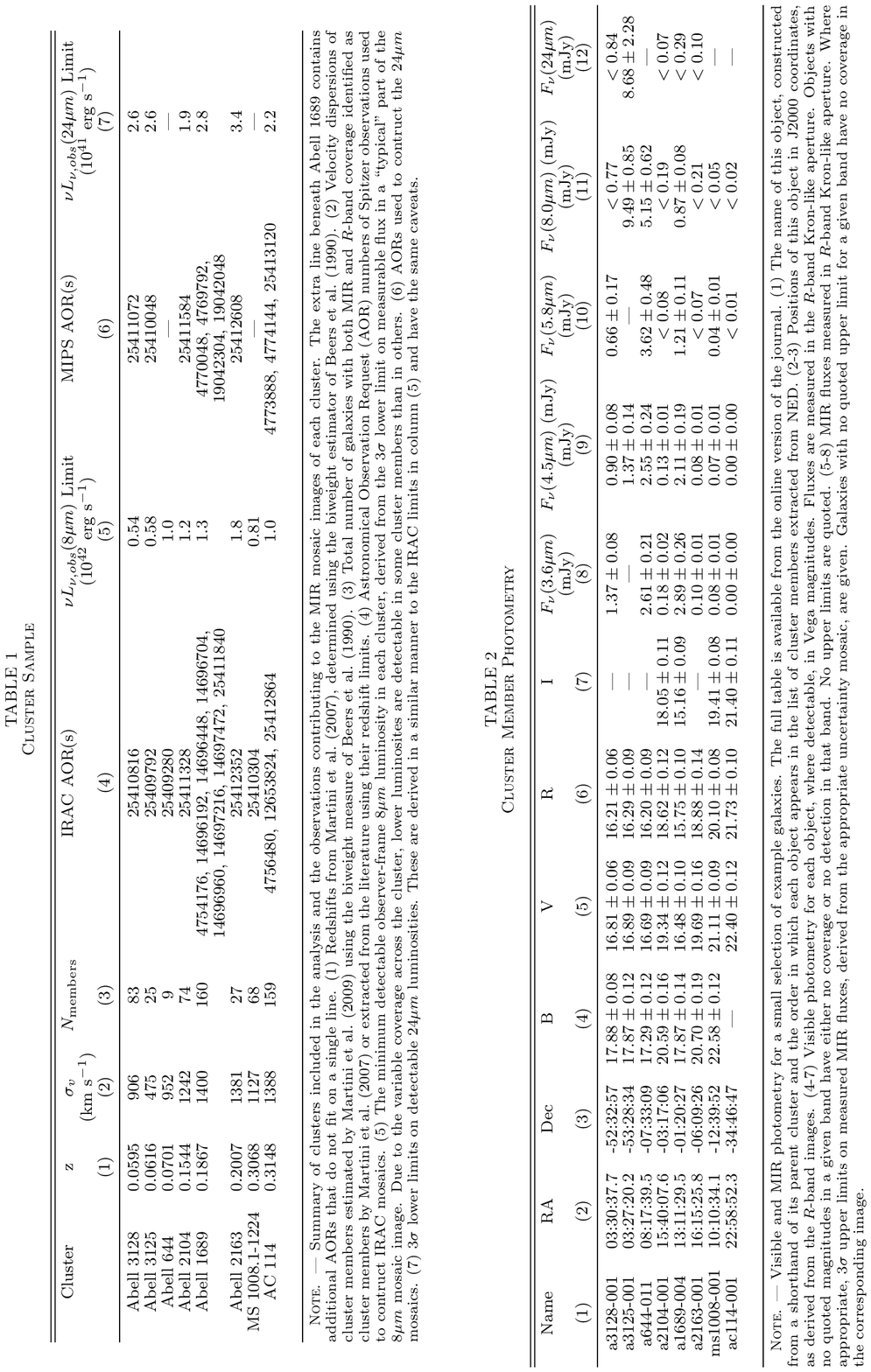}
\end{figure*}

\clearpage
\input{tab3.tex}

\input{tab4.tex}

\end{document}

%% file: tab3.tex
\begin{deluxetable}{cccccccc}
\tabletypesize{\scriptsize}
\tablewidth{0pc}
\tablecaption{Identified Active Galactic Nuclei
\label{tabAgn}}
\tablehead{\colhead{Name} & \colhead{Martini Name} & \colhead{RA} &
\colhead{Dec} & \colhead{$L_{bol} (10^{43}erg\ s^{-1})$} &
\colhead{$L_{\rm X}$ $(10^{41}erg\ s^{-1})$} & 
\colhead{$M_{*} (10^{10}M_{\odot})$} & \colhead{SFR $(M_{\odot}\ yr^{-1})$} \\
\colhead{(1)} & \colhead{(2)} & \colhead{(3)} & \colhead{(4)} & \colhead{(5)} & 
\colhead{(6)} & \colhead{(7)} & \colhead{(8)}}
\startdata
a3128-004 & a3128-9 & 03:30:39.3 & -52:32:05 & $  1.5\pm  0.5$ & $  3.8$ & $  3.2\pm  0.9({\rm stat})\pm  1.9({\rm syst})$ & $< 0.04$ \\
a3128-012 & a3128-6 & 03:30:17.3 & -52:34:08 & $  1.0\pm  0.3$ & $ 22.4$ & $  1.9\pm  0.6({\rm stat})\pm  1.1({\rm syst})$ & $< 0.35$ \\
a3128-092 & a3128-2 & 03:29:41.4 & -52:29:35 & --- & $ 22.4$ & $  1.1\pm  0.2({\rm stat})\pm  0.6({\rm syst})$ & $< 0.44$ \\
a3125-044 & a3125-5 & 03:27:05.0 & -53:21:41 & --- & $  7.2$ & $  7.5\pm  1.6({\rm stat})\pm  4.5({\rm syst})$ & $  2.7\pm  0.4$ \\
a644-011 & a644-1 & 08:17:39.5 & -07:33:09 & $ 10.4\pm  1.4$ & $ 28.2$ & $  2.6\pm  0.9({\rm stat})\pm  1.6({\rm syst})$ & $< 1.38$ \\
a644-024 & a644-2 & 08:17:48.1 & -07:37:31 & --- & $  4.5$ & $  7.6\pm  1.8({\rm stat})\pm  4.5({\rm syst})$ & $  1.2\pm  0.4$ \\
a2104-024 & a2104-4 & 15:40:14.0 & -03:17:03 & --- & $  7.2$ & $  2.0\pm  0.6({\rm stat})\pm  1.2({\rm syst})$ & $< 0.50$ \\
a2104-040 & a2104-6 & 15:40:03.9 & -03:20:38 & --- & $  7.2$ & $ 12.4\pm  3.8({\rm stat})\pm  7.4({\rm syst})$ & $  1.9\pm  0.6$ \\
a2104-046 & a2104-5 & 15:40:19.5 & -03:18:24 & --- & $ 36.3$ & $  1.7\pm  0.5({\rm stat})\pm  1.0({\rm syst})$ & $< 0.83$ \\
a2104-051 & a2104-2 & 15:40:16.7 & -03:15:07 & $  4.1\pm  0.8$ & $ 18.2$ & $<  2.3$ & $<16300.00$ \\
\enddata
\tablecomments{Brief sample table summarizing AGNs identified either by their
X-ray luminosity or their SED shapes.  The full table is available from the
electronic edition of the journal.  
(1) The name of this object in Table \ref{tabPhot}.  (2) The name given to
the X-ray source by \citet{mart06}.
(3-4) Position of this
AGN in J2000 coordinates, as derived from the $R$-band image.
(5) The bolometric luminosity derived by integrating
the direct component of the AGN contribution to the model SED.
These luminosities are quoted
only for IR AGNs.  (6) Rest-frame X-ray luminosities in the 0.3-8 keV
band from Table 4 of \citet{mart06}.  X-ray luminosities are given only
for X-ray AGNs.
(7) Stellar mass derived using the M/L coefficients appropriate for
a solar metallicity galaxy with a scaled Salpeter IMF and applying the
Bruzual \& Charlot population synthesis model (\citealt{bell01}, Table
4).  Systematic errors are derived by applying the M/L coefficients appropriate
for a Salpeter IMF and the {\sc{P\'egase}} population synthesis model.
Upper limits are given at $3\sigma$ of the statistical error only.
(8) SFR derived either from the $8\mu m$ luminosity, the $24\mu m$
luminosity or by taking the geometric mean of the two, depending on
the measurements available.  Uncertainties include only statistical
errors, and upper limits are quoted at $3\sigma$ in the more sensitive
of the $8\mu m$ and $24\mu m$ bands.}
\end{deluxetable}

%% file: tab4.tex
\begin{deluxetable}{rrcccccc}
\tabletypesize{\scriptsize}
\tablewidth{0pc}
\tablecaption{IR AGN Selection Efficiency
\label{tabComplete}}
\tablehead{
\colhead{} & \colhead{${\bf L_{AGN}:}$} & 
\colhead{${\bf -100 \rightarrow -1.0}$} & \colhead{${\bf -1.0 \rightarrow -0.4}$} & 
\colhead{${\bf -0.4 \rightarrow 0.2}$} & \colhead{${\bf 0.2 \rightarrow 0.8}$} &
\colhead{${\bf 0.8 \rightarrow 1.5}$} & \colhead{${\bf 1.5 \rightarrow 100}$} \\
\colhead{${\bf E(B-V)}$} & \colhead{} & \colhead{} & \colhead{} & \colhead{} &
\colhead{} & \colhead{} & \colhead{}}
\startdata
{\bf 0--0.075} & & $6.9\pm0.3\%$ & $8.1\pm0.3\%$ & $17.8\pm0.4\%$ & $45.5\pm0.9\%$ & $72.9\pm1.9\%$ & $83.8\pm4.6\%$ \\
{\bf 0.075--0.15} & & $7.0\pm0.4\%$ & $8.3\pm0.3\%$ & $17.8\pm0.4\%$ & $45.6\pm0.6\%$ & $75.5\pm2.0\%\%$ & $83.8\pm5.0\%$ \\
{\bf 0.15--0.3} & & $7.0\pm0.3\%$ & $8.3\pm0.3\%$ & $17.1\pm0.4\%$ & $44.7\pm0.8\%$ & $74.2\pm1.7\%$ & $84.3\pm4.2\%$ \\
{\bf 0.3--0.4} & & $6.7\pm0.5\%$ & $7.5\pm0.4\%$ & $18.2\pm0.6\%$ & $46.2\pm1.3\%$ & $77.2\pm2.7\%$ & $86.8\pm6.9\%$ \\
{\bf 0.4--0.6} & & $7.2\pm0.4\%$ & $8.4\pm0.3$ & $17.8\pm0.5\%$ & $49.2\pm1.0\%$ & $77.0\pm2.1\%$ & $85.8\pm5.1\%$ \\
{\bf 0.6--1.0} & & $7.2\pm0.4\%$ & $8.4\pm0.3\%$ & $17.8\pm0.5\%$ & $49.2\pm1.0\%$ & $77.0\pm2.1\%$ & $85.8\pm5.1\%$ \\
{\bf 1.0--2.0} & & $6.7\pm0.5\%$ & $7.9\pm0.3\%$ & $17.4\pm0.5\%$ & $47.8\pm1.1\%$ & $76.2\pm2.2\%$ & $84.1\pm5.6\%$ \\
{\bf 2.0--100} & & $6.8\pm0.7\%$ & $7.1\pm0.6\%$ & $14.2\pm0.7\%$ & $37.2\pm1.6\%$ & $65.4\pm3.4\%$ & $83.1\pm9.9\%$ \\
\enddata
\tablecomments{Breakdown of AGN selection efficiency
by $\log[L_{AGN}/10^{10}L_{\odot}]$ and $E(B-V)$.  Efficiencies were
determined using Monte Carlo to generate model SEDs with varying
contributions from the three normal galaxy templates, add
AGN components to the model SED,
and introduce photometric errors.
See \S~\ref{secAgn} for further details.}
\end{deluxetable}